\documentclass[12pt]{article}
\usepackage{mathrsfs}
\usepackage{mathrsfs}
\usepackage{amssymb}
\usepackage{color}
\usepackage{caption}
\usepackage{psfrag}
\usepackage[center]{subfigure}
\usepackage{rotating}
\usepackage{epstopdf}
\usepackage{mathrsfs}
\usepackage{latexsym}
\usepackage[numbers,sort&compress]{natbib}
\newcommand{\bea}{\begin{eqnarray}}
\newcommand{\eea}{\end{eqnarray}}
\newcommand{\be}{\begin{equation}}
\newcommand{\ee}{\end{equation}}
\newcommand{\vs}[1]{\vspace{#1 mm}}

\newcommand{\dsl}{\pa \kern-0.5em /}

\newcommand{\pa}{\partial}

\newcommand{\nn}{\nonumber\\}

\newcommand{\eqn}[1]{(\ref{#1})}

\begin{document}
\topmargin 0mm
\oddsidemargin 0mm

\begin{flushright}


\end{flushright}

\vspace{2mm}

\begin{center}

{\Large \bf Modification of phase structure of black $D6$
branes in a canonical ensemble and its origin}

\vs{10}

{\large J. X. Lu$^{a,}$\footnote{jxlu@ustc.edu.cn},
Jun Ouyang,$^{a,}$\footnote{yangjun@mail.ustc.edu.cn} and Shibaji
Roy$^{b,}$\footnote{shibaji.roy@saha.ac.in}}

\vspace{4mm}

{\em

 $^a$ Interdisciplinary Center for Theoretical Study\\
 University of Science and Technology of China, Hefei, Anhui
 230026, People's Republic of China\\

\vs{4}

 $^b$ Saha Institute of Nuclear Physics,
 1/AF Bidhannagar, Calcutta-700 064, India\\}

\end{center}

\vs{10}

\begin{abstract}
It is well known that charged black $Dp$ branes of type II string theory share
a universal phase structure of van der Waals$-$Maxwell liquid-gas type except
$D5$ and $D6$ branes. Interestingly, the phase structure of $D5$ and $D6$ branes can
be changed to the universal form with the inclusion of particular delocalized
charged lower-dimensional branes. For $D5$ branes one needs to introduce delocalized
$D1$ branes, and for $D6$ branes, one needs to introduce delocalized $D0$ branes to
obtain the universal structure. In a previous paper [J. High Energy Phys. {\bf 04} (2013) 100]
Lu with Wei study the phase structure of black $D6$ branes
with the introduction of delocalized $D0$ branes in a special case when their
charges are equal and the dilaton charge vanishes. In this paper, we look at
the phase structure of the black $D6/D0$ system with the generic values of the parameters,
which makes the analysis more involved but the structure more rich.
We also provide reasons why the respective modifications of the phase structures to the
universal form for the black $D5$ and $D6$ branes occur when specific delocalized
lower-dimensional branes are introduced.
\end{abstract}

\newpage

\section{Introduction}

It is quite well known that charged AdS black holes give rise to an interesting thermodynamic phase structure isomorphic to the
van der Waals$-$Maxwell liquid-gas system\cite{Chamblin:1999tk, Chamblin:1999hg} (See, also,
\cite{Kubiznak:2012wp, Gunasekaran:2012dq, Altamirano:2013uqa, Smailagic:2012cu, Spallucci:2013osa, Spallucci:2013jja, Nicolini:2011dp}
for some recent discussions on related issues.). The interest
in AdS black hole stems from the fact that they are thermodynamically stable and, hence, are
suitable to study equilibrium thermodynamics\cite{Hawking:1982dh}. Moreover, by AdS/CFT correspondence, they are
holographically dual to finite temperature field theories\cite{Witten:1998zw}, and, indeed, the above-mentioned phase
structure in the field theory has similarities with catastrophe theory\cite{Chamblin:1999hg}. However, it has been noted before
that the large part of this phase structure including the van der Waals$-$Maxwell liquid-gas type
is not unique to the AdS black holes only but appear in suitably stabilized dS as well as asymptotically flat
space charged black holes \cite{Carlip:2003ne, Lundgren:2006kt}. Such universal structure for the charged black holes with different
asymptopia suggests that holography might be at work not just for AdS space, but for dS as well as
flat space\cite{Carlip:2003ne,Lundgren:2006kt, Lu:2010xt, Lu:2012rm, Lu:2013nt}\footnote{However, the study of the
present $D6/D0$ system, in particular, indicates that a universal thermodynamical phase structure of the underlying gravity system
merely reflects its interesting thermal properties and does not necessarily imply a holography. We discuss this in detail in Sec.4.}

Motivated by this, in \cite{Lu:2010xt} the phase structure of suitably stabilized, flat,
charged black $p$ brane solutions in arbitrary dimensions was analyzed, and, surprisingly, it was found that
they also have very similar phase structure as that of the black holes and, in particular, they have the van der Waals$-$Maxwell
liquid-gas$-$type structure when the charge of the black brane is below a certain nonzero critical value.
However, this happens only when the dimensions of the space transverse to the $p$ brane satisfy $\tilde{d}+2 =
D-p-1 > 4$, where $D$ is the total space-time dimension. When $D=10$, i.e., for string theory branes,
this implies that all the charged black $Dp$ branes with $p<5$ share the same universal phase structure
as the charged black holes in AdS/dS/flat space, but the phase structures of $D5$ and $D6$ branes differ. It
was found in \cite{Lu:2012rm, Lu:2013nt}, that this difference in phase structure can be removed if one adds specific delocalized
charged lower-dimensional branes to the system. So, for example, $D5$ branes restore the same universal phase structure
if one adds delocalized $D1$ branes to the system; on the other hand, $D6$ branes restore the universal phase
structure if one adds delocalized $D0$ branes to the system. Note that for $D5$ branes, adding other lower-dimensional
branes, namely, the delocalized charged $D3$ branes, does not help produce the universal phase structure. Similarly, for
$D6$ branes, the other charged lower-dimensional branes, namely, the delocalized $D4$ or $D2$ branes, do not help even though
the $D6/D2$ system belongs to the same class of $Dp/D(p - 4)$ as the $D5/D1$ system. This shows that in order to obtain the universal
phase structure from $D5$ or $D6$ branes, it is not {\it a priori} clear which charged lower-dimensional branes one should
include to the system if they can bring about this change at all. Also, it should be emphasized that the inclusion
of lower-dimensional branes does not automatically imply that they will make the necessary change in phase structure,
as one might think, since the lower-($<5$) dimensional branes themselves have universal phase structure. In
fact, one can check that the delocalized (in the other four  $D5$ world-volume spatial directions) charged $D1$ branes and delocalized (in the six  $D6$ world-volume
spatial directions) charged $D0$ branes individually have the same phase structure as $D5$ branes and $D6$ branes, respectively\cite{Lu:2012rm, Lu:2013nt}. Thus, when
they are combined to form bound states, it is their interactions with each other which makes the necessary change in the phase structure possible.

The phase structure of the black $D5/D1$ system has been studied with all its generalities in \cite{Lu:2012rm}. The black $D6/D0$
system with the associated phase structure has been studied in \cite{Lu:2013nt}. As the parameter space of the latter system
is quite complicated to analyze in general, only a special case has been considered which enabled the authors to show the
universal phase structure, postponing the discussion of the general case as well as the reason behind the appearance of
this universal structure to a later publication. It is this task that we undertake in this paper. In the previous
publication, the charges of the $D6$ branes and the $D0$ branes were chosen to be equal, which gives the vanishing dilaton charge from the parameter
relation.\footnote{The other two solutions each give a naked singularities and, therefore, are not relevant for thermodynamical consideration.}
 However, in this paper we look at the general case, which makes the
solution of the parameter space far more complicated, and the phase structure, which has the expected universal form,
becomes richer than before. We also provide possible reasons why the additions of delocalized $D1$ branes in $D5$ branes
and $D0$ branes in $D6$ branes change qualitatively the thermodynamic phase structure of $D5$ and $D6$ branes to have
the universal form. In the former case, it is the addition of extra degrees of freedom or the change in entropy, while in
the latter case, it is the repulsive nature of interaction between the constituent branes which makes the necessary
change in the thermodynamic phase structure to take the universal form.

This paper is organized as follows. In Sec. 2, we discuss the general charged black $D6/D0$ bound state solution
in Euclidean signature and describe the general parameter space for which there exists a regular horizon such that
a meaningful thermodynamics can be given. The corresponding thermodynamics and the phase structure are described in Sec. 3. We provide reasons for the appearance of the universal phase structure of van der Waals$-$Maxwell liquid-gas
type in Sec. 4. Finally, we give our concluding remarks in Sec. 5.

\section{$D6/D0$ bound state and the parameter space}

In this section, we write the spherically symmetric, time-independent, electrically charged black $D6/D0$ bound state solution
in Euclidean signature for the purpose of studying thermodynamics and phase structure\cite{Brandhuber:1997tt, Dhar:1998ip}. As we will see, the solution contains
three independent parameters: the mass and the charges of $D6$ branes and $D0$ branes. We will argue that the parameters
cannot take arbitrary values, as naked singularities can develop in general. We will determine the region of the parameter space
for which there exists a regular horizon. The $D6/D0$ solution is given as\footnote{Here we use the configuration given
in \cite{Brandhuber:1997tt} with some modifications. The magnetic part of the 1-form given there has been changed to an electric 7-form. In
$D=10$, $D0$ branes and $D6$ branes are electric-magnetic dual to each other. Also, we change the sign of the charge parameters $Q$ and
$P$ there and assume without any loss of generality $Q > 0$, $P > 0$ for convenience. Further, we correct a typo in the electric
1-form potential given there by replacing the dilaton charge $\Sigma$ by $\Sigma/\sqrt{3}$.}
\bea \label{bd0d6}
ds^{2}
&=& F A^{-\frac{1}{8}} B^{-\frac{7}{8}} dt^{2}  + \left(B /
A\right)^{\frac{1}{8}}\sum_{i = 1}^{6} dx_i^{2} +
{A}^{\frac{7}{8}}{B}^{\frac{1}{8}} \left( F^{- 1} d \rho^{2} +
\rho^{2} d\Omega_{2}^{2} \right)\nn A_{[1]}&=&i e^{- 3 \phi_{0}/4}\,
Q \left[\frac{1 - \frac{\Sigma}{\rho_+ \sqrt{3}}}{\rho_+ B (\rho_+)}
- \frac{1 - \frac{\Sigma}{\rho \sqrt{3}}}{\rho B (\rho)}\right]dt\nn
A_{[7]}&=&i e^{3 \phi_{0}/4} \, P \left[\frac{1 +
\frac{\Sigma}{\rho_+ \sqrt{3}}}{\rho_+ A (\rho_+)}- \frac{1 +
\frac{\Sigma}{\rho \sqrt{3}}}{\rho A (\rho)}\right]dt\wedge
dx^{1}...\wedge dx^{6}\nn e^{2(\phi-\phi_{0})}&=& \left(B (\rho)/
A(\rho)\right)^{3/2},
\eea
where the metric in \eqn{bd0d6} is given in the Einstein frame, and the various functions appearing in the metric
are defined as
\bea \label{fab}
F (\rho) &=& \left(1 - \frac{\rho_+}{\rho}\right)\left(1 -
\frac{\rho_-}{\rho}\right),\nn A (\rho) &=& \left(1 -
\frac{\rho_{A+}}{\rho}\right) \left(1 -
\frac{\rho_{A-}}{\rho}\right), \nn B (\rho) &=& \left(1 -
\frac{\rho_{B+}}{\rho}\right) \left(1 -
\frac{\rho_{B-}}{\rho}\right),\eea
with
\bea \label{rho-pm}
\rho_\pm &=& M \pm \sqrt{M^2 + \Sigma^2 - P^2/4 - Q^2/4},\nn \rho_{A\pm} &=&
\frac{\Sigma}{\sqrt{3}} \pm \sqrt{\frac{P^2 \Sigma/2}{\Sigma -
\sqrt{3} M}},\nn \rho_{B\pm} &=& - \frac{\Sigma}{\sqrt{3}} \pm
\sqrt{\frac{Q^2 \Sigma/2}{\Sigma + \sqrt{3} M}}.\eea
In \eqn{rho-pm}, there are four parameters, namely, the mass parameter $M$, the delocalized $D0$ brane
charge parameter $Q$, the $D6$ brane charge parameter $P$, and the dilaton charge parameter $\Sigma$.
However, not all of them are independent, and, in fact, dilaton charge parameter $\Sigma$ is related to
$M$, $P$, and $Q$ by the relation
\be \label{dilatoncharge}
\frac{8}{3}\Sigma = \frac{Q^2}{\Sigma + \sqrt{3} M} +
\frac{P^2}{\Sigma - \sqrt{3} M},\ee
leaving only three of them independent. As noted in \cite{Brandhuber:1997tt}, under the electric-magnetic duality, the parameters
of the solution transform as $Q \leftrightarrow P$, $\Sigma \leftrightarrow - \Sigma$, and $M\leftrightarrow M$.
Also, in \eqn{bd0d6}, $A_{[1]}$ and $A_{[7]}$ are the electric 1-form
and 7-form to which $D0$ branes and $D6$ branes couple and give the corresponding charges $Q$ and $P$, respectively.
The form fields are chosen to vanish at $\rho_+$ so that they are well defined in the local inertial frame,
and $\phi_0$ is the asymptotic value of the dilaton.

Note that the solution \eqn{bd0d6} given in terms of three parameters $M$, $Q$, and $P$ is not necessarily physical
as for generic values of these parameters it can have naked singularity. We will see in this section that for some
restricted region of the parameter space, we can, indeed, have a physical solution with a well-defined horizon, which, in turn, will
be suitable for studying thermodynamics and the associated phase structure. Also, in addition to $P,\,Q>0$, we will
assume by duality symmetry that $\Sigma \geq 0$ without loss of generality. The $\Sigma < 0$ branch can be obtained from
$\Sigma > 0$ simply by exchanging $Q \leftrightarrow P$. The three quantities, which will be useful for showing the existence of a regular
horizon, are $\rho_+$, $\rho_{A+}$, and $\rho_{B+}$ and are given in terms of the parameters of the solution as
\bea \label{threeq}
   \rho_+
&=& M + \sqrt{M^2 + \Sigma^2 - P^2/4 - Q^2/4},\nn \rho_{A +} &=&
\frac{\Sigma}{\sqrt{3}} + \sqrt{\frac{P^2 \Sigma/2}{\Sigma -
\sqrt{3} M}},\nn \rho_{B +} &=& - \frac{\Sigma}{\sqrt{3}} +
\sqrt{\frac{Q^2 \Sigma/2}{\Sigma + \sqrt{3} M}}. \eea
Actually, $\rho =\rho_+$ is the horizon as long as it is greater than both $\rho_{A+}$ and $\rho_{B+}$.\footnote{Note that the metric \eqn{bd0d6} has curvature singularities at
both $\rho = \rho_{A+}$ and $\rho = \rho_{B+}$.} Let us first assume
that $\Sigma \geq \sqrt{3}M$. Now it can be easily checked from \eqn{threeq} that with this $\rho_{A+} > \rho_{B+}$. So,
in order to get a horizon at $\rho_+$, we must have $\rho_+ > \rho_{A+}$. However, using their expressions from \eqn{threeq},
we find that this condition cannot be satisfied. Thus, if $\Sigma \geq \sqrt{3}M$, the solution has a naked singularity at
$\rho = \rho_{A+}$. Therefore, in order to have a horizon (if it exists at all), we must take $0 < \Sigma < \sqrt{3}M$. Note that we have
excluded the $\Sigma =0$ case since it corresponds to [from \eqn{dilatoncharge}] $Q^2 = P^2$, which has been considered in \cite{Lu:2013nt}.
Now, as $\Sigma < \sqrt{3}M$, we can see from \eqn{rho-pm} that $\rho_{A\pm}$ are both imaginary and, therefore, do not play any role
in determining whether there exists a horizon. We, therefore, must demand $\rho_+ > \rho_{B+}$ in order to have a well-defined horizon.
Note that $\rho_{B+} > 0$, which can be verified using \eqn{dilatoncharge}. Using the form of $\rho_+$ and $\rho_{B+}$ from \eqn{threeq}
and after some algebraic manipulation and further using \eqn{dilatoncharge}, the condition $\rho_+ > \rho_{B+}$ gives
\be\label{rhorhoB} \sqrt{M^2 + \Sigma^2 - P^2/4 - Q^2/4} > \frac{\sqrt{3}}{4} \frac{P^2}{\sqrt{3} M - \Sigma} -
\frac{1}{\sqrt{3}}\left(\sqrt{3} M - \Sigma\right).\ee
It can be easily checked that if the rhs of \eqn{rhorhoB} is positive, i.e., if $\sqrt{3}M - \Sigma \leq \frac{\sqrt{3}}{2} P$, then
\eqn{rhorhoB} implies $\Sigma < 0$ when \eqn{dilatoncharge} is used. This is a contradiction to our assumption that $\Sigma > 0$. Therefore, we must have
$\sqrt{3}M - \Sigma > \frac{\sqrt{3}}{2} P$, or, in other words, the rhs of \eqn{rhorhoB} must be negative. So, to summarize,
in order to have a well-defined horizon, we must have at least
\be\label{sigmacond}
\Sigma > 0, \qquad {\rm and} \qquad \sqrt{3}M - \Sigma > \frac{\sqrt{3}}{2} P \quad \Rightarrow \quad \sqrt{3}M > \Sigma +
\frac{\sqrt{3}}{2}P
\ee
along with \eqn{dilatoncharge}.

We will see that the condition \eqn{dilatoncharge} and the positivity of the quantity inside the square root of the expression
of $\rho_+$ given in \eqn{threeq} will put more restrictions on $\Sigma$ in terms of $P$ and $Q$ in order to have well-defined
horizon. Let us first look at the condition \eqn{dilatoncharge}. Defining $X=\sqrt{3}M > 0$, we rewrite it as
\be\label{xeq}
X^2 + \frac{3 (P^2 - Q^2)}{8 \Sigma} X + \frac{3}{8} (Q^2 + P^2) - \Sigma^2 = 0,\ee
from which we solve $X$ to get
\be\label{xpm}
X_\pm = \frac{3 (Q^2 - P^2)}{16 \Sigma} \pm \frac{1}{2}\sqrt{\frac{9 (Q^2 - P^2)^2}{64 \Sigma^2} + 4 \Sigma^2 - \frac{3}{2}(Q^2 + P^2)}.\ee
Let us now check the condition \eqn{sigmacond}, i.e., $X_{+} > \Sigma + (\sqrt{3}/2) P$. Using $X_{+}$ given in \eqn{xpm} we get
\bea\label{cond1} & & \frac{3 (Q^2 - P^2)}{8 \Sigma} + \sqrt{\frac{9 (Q^2 - P^2)^2}{64 \Sigma^2} + 4 \Sigma^2 -
\frac{3}{2}(Q^2 + P^2)} > 2 \Sigma + \sqrt{3}P,\nn
{\rm or} & & \frac{9 (Q^2 - P^2)^2}{64 \Sigma^2} + 4 \Sigma^2 - \frac{3}{2}(Q^2 + P^2) > \left(2 \Sigma + \sqrt{3}P -
\frac{3 (Q^2 - P^2)}{8 \Sigma}\right)^2.
\eea
In writing the second inequality in \eqn{cond1}, we have assumed $2\Sigma + \sqrt{3}P \geq 3(Q^2-P^2)/(8\Sigma)$, which is certainly true
if $Q^2 < P^2$. However, we note that the second inequality in \eqn{cond1} leads to a contradiction since it gives $2 \Sigma + \sqrt{3} P <
3(Q^2 - P^2)/(8\Sigma)$. Therefore, we must have
$2\Sigma + \sqrt{3}P < 3(Q^2-P^2)/(8\Sigma)$. This not only implies $Q > P$ but also ensures that the quantity inside the square root of the
expression of $X_{\pm}$ given in \eqn{xpm} is positive definite. From this condition, we have
\be\label{cond2} \left(\Sigma + \frac{\sqrt{3}}{4} (Q + P)\right)\left(\Sigma - \frac{\sqrt{3}}{4} (Q - P)\right) < 0,\ee
which gives a restriction on $\Sigma$ as
\be \label{cond3} \Sigma < \frac{\sqrt{3}}{4} (Q - P).
\ee
For $X_{-}$, it can be easily checked that the condition $X_- > \Sigma + (\sqrt{3}/2)P$ is contradictory with $2\Sigma + \sqrt{3}P < 3(Q^2-P^2)/(8\Sigma)$,
and, therefore, $X_-$ is not a valid solution for our discussion. In summary, so far we find that the solution \eqn{bd0d6} has a horizon, i.e., $\rho_+ > \rho_{B+}$, if
\bea \label{cond4}
&&\sqrt{3} M = \frac{3 (Q^2 - P^2)}{16 \Sigma} + \frac{1}{2}\sqrt{\frac{9 (Q^2 - P^2)^2}{64 \Sigma^2} + 4 \Sigma^2 -
\frac{3}{2}(Q^2 + P^2)},\nn
&& 0 < \Sigma < \frac{\sqrt{3}}{4} (Q - P),\,\,\,\quad  Q > P.
\eea

We may think that we have fixed the parameter space, but this is not quite true. We have to consider, as we mentioned before, one more condition
coming from the quantity inside the square root of $\rho_+$ given in \eqn{threeq} which must be positive semidefinite.

Therefore, from the expression of $\rho_+$ given in \eqn{threeq}, we have
\be\label{rhoplus1}
M^2 \geq \frac{P^2 + Q^2}{4} - \Sigma^2.
\ee
Using the expression for $M$ given in \eqn{cond4}, the above relation reduces to
\be \label{newcond1}\frac{3 (Q^2 - P^2)}{16 \Sigma} + \frac{1}{2}\sqrt{\frac{9 (Q^2 - P^2)^2}{64 \Sigma^2} + 4 \Sigma^2 -
\frac{3}{2}(Q^2 + P^2)} \geq \sqrt{3} \sqrt{\frac{P^2 + Q^2}{4} - \Sigma^2}.\ee
From the above, it is clear that \eqn{newcond1} will be automatically satisfied if we have
\be\label{newcond2}
\frac{\sqrt{3}}{16} \frac{Q^2 - P^2}{\Sigma} \geq \sqrt{\frac{P^2 + Q^2}{4} - \Sigma^2}.
\ee
From this, we will determine the condition for $\Sigma$. Equation \eqn{newcond2} can be simplified as
\be\label{newcond3}
\Sigma^4 - \frac{P^2 + Q^2}{4} \Sigma^2 + \frac{3(Q^2 - P^2)^2}{16^2} \geq 0,\ee
which gives
\be\label{newcond4} (\Sigma^2 - \Sigma_+^2)(\Sigma^2 - \Sigma_-^2) \geq 0,\ee
where
\be \label{sigmazf} \Sigma_\pm^2 = \frac{P^2 + Q^2}{8} \pm \frac{1}{8} \sqrt{\frac{P^4 + Q^4 + 14 P^2 Q^2}{4}}.\ee
From \eqn{newcond4}, we have either $\Sigma^2 \geq \Sigma_+^2$ or $\Sigma^2 \leq \Sigma_-^2$. We also need $\Sigma
< \sqrt{3}(Q-P)/4$ from \eqn{cond4}. But it can be easily shown that $\Sigma_+ > \sqrt{3}(Q-P)/4$ and so it is not
relevant; however, $\Sigma_- < \sqrt{3}(Q-P)/4$, and, therefore, $\Sigma_-$ sets a new bound on $\Sigma$, i.e., $\Sigma
< \Sigma_-$.

However, this is not the complete story. We still need to consider the case
\be\label{newcond5}
\frac{\sqrt{3}}{16} \frac{Q^2 - P^2}{\Sigma} < \sqrt{\frac{P^2 + Q^2}{4} - \Sigma^2},
\ee
such that the inequality \eqn{newcond1} holds. This can give further restrictions on $\Sigma$. We rewrite \eqn{newcond1} as
\be\label{newcond6} \sqrt{\frac{9 (Q^2 - P^2)^2}{64 \Sigma^2} + 4 \Sigma^2 - \frac{3}{2}(Q^2 + P^2)} \geq 2 \sqrt{3} \sqrt{\frac{P^2 + Q^2}{4}
- \Sigma^2} - \frac{3 (Q^2 - P^2)}{8 \Sigma} > 0.
\ee
Now, squaring both sides and doing some algebraic manipulations, we get
\be\label{newcond7}
\frac{9}{32}(Q^2+P^2) - \Sigma^2 \leq \frac{3 \sqrt{3} (Q^2 - P^2)}{32 \Sigma} \sqrt{ \frac{Q^2 + P^2}{4} - \Sigma^2}.
\ee
Now, let us define a dimensionless variable $y=4\Sigma^2/(P^2+Q^2)$, then in terms of $y$, \eqn{newcond7} can be rewritten as
\be\label{ycond}
y^3 - \frac{9}{4} y^2 + \frac{81+27k^2}{64}y - \frac{27k^2}{64} \leq 0,
\ee
where
\be\label{a}
k = \frac{Q^2-P^2}{Q^2+P^2} < 1.
\ee
The left side of the above inequality \eqn{ycond} can actually be factorized, and it can be written as
\be \left[y - \frac{3}{4}\left(1 - \frac{A}{2}\right)\right]\left[\left(y - \frac{3}{4}\left(1 + \frac{A}{4}\right)
\right)^2 + \frac{3^3}{16^2}(1 - k^2)^{\frac{2}{3}}\left((1 + k)^{\frac{1}{3}} - (1 - k)^{\frac{1}{3}}\right)^2\right] \leq 0,\ee
 where
\be A = (1 - k^2)^{1/3}[(1 + k)^{1/3} + (1 - k)^{1/3}] < 2.\ee We, therefore, have
\be y \leq \frac{3}{4}\left(1 - \frac{A}{2}\right),\ee which gives, after plugging the definition for $y$,
\be \label{s0} \Sigma \leq \Sigma_0 = \frac{\sqrt{3}}{4} (Q - P)\left[1 - \frac{(QP)^{2/3}}{(Q^{2/3} + P^{2/3} + Q^{1/3} P^{1/3})^2}\right]^{1/2}
< \frac{\sqrt{3}}{4} (Q - P).\ee
Now, in order to show that $\Sigma_0$ is the correct bound, we need to show $\Sigma_0 > \Sigma_-$, where $\Sigma_-$ is given in
\eqn{sigmazf}. For this, let us compare the expressions for $\Sigma_0^2$ from \eqn{s0} and $\Sigma_-^2$ from \eqn{sigmazf}. They
have the forms,
\bea\label{sminuss0}
\Sigma_0^2 &=& \frac{3}{16}\left[(Q-P)^2 - (QP)^{2/3}\left(Q^{1/3} - P^{1/3}\right)^2\right],\nn
\Sigma_-^2 &=& \frac{Q^2+P^2}{8} - \frac{1}{16}\sqrt{P^4 + Q^4 + 14 P^2Q^2} < \frac{(Q-P)^2}{8}
\eea
So, all we need to show is $\Sigma_0^2 > (Q-P)^2/8$. Now, substituting the form of $\Sigma_0$ from \eqn{s0}, this condition
leads to
\be\label{s0geqsm}
\left(Q^{1/3} - P^{1/3}\right)^2 + 3 (QP)^{1/3} > \sqrt{3}(QP)^{1/3},
\ee
which obviously holds true, and, therefore, this shows that $\Sigma_0 > \Sigma_-$.

So, finally we conclude that for the existence of a sensible horizon for the $D6/D0$ black brane bound state
solution, we must have
\be\label{finalcond}
\sqrt{3}M =  \frac{3 (Q^2 - P^2)}{16 \Sigma} + \frac{1}{2}\sqrt{\frac{9 (Q^2 - P^2)^2}{64 \Sigma^2} + 4 \Sigma^2 - \frac{3}{2}(Q^2 + P^2)},\ee with
  \be \label{fc1} 0 < \Sigma \leq \Sigma_0,\ee
  and $Q > P$, where $\Sigma_0$ is given in \eqn{s0}. We can extend the above range of $\Sigma$ to $- \Sigma_0 \leq \Sigma < 0$ if we require
$P > Q$. For our purpose, we will focus on the $Q > P$ branch in what follows.

\section{The general phase structure of $D6/D0$}

In this section, we will analyze the phase structure of the black $D6/D0$ system with generic charges with the parameters $M$ and
$\Sigma$ satisfying the condition given in \eqn{finalcond} and \eqn{fc1} for the existence of a well-defined horizon.  Since this system
 is asymptotically flat, we need to stabilize it by placing it in a cavity following \cite{York:1986it,Lu:2010xt}, and in this paper we will analyze the
phase structure in a canonical ensemble which will be specified later on. All we need to know is the form of the local temperature (or the inverse of the
local temperature to be precise) of the system at the location of the wall of the cavity,  which can be obtained from the black $D6/D0$ metric in Euclidean
signature as given in \eqn{bd0d6} by demanding the absence of conical singularity at the horizon.
We will express the inverse of the local temperature at the given location as a function of the
horizon radius only, and, therefore, we need to express the other parameters, namely, $M$ and $\Sigma$, also in terms of the horizon
radius. However, for this system and from our past experience \cite{Lu:2013nt}, we know that $\rho$ is not a good coordinate for this purpose, and we will define
a new radial coordinate by
\be\label{newcoord}
r = \rho + a,
\ee
where $a$ is a parameter to be determined later. From now on, we will assume $Q>P$ and $0 < \Sigma \leq \Sigma_0$. From
\eqn{newcoord} and using the new radial coordinate $r$, we have $r_+ = \rho_+ + a$ and $r_- = \rho_- + a$, where $r_+$ defines the location of horizon, and using these two we have
\be\label{rplusrminus}
r_+ r_- = \rho_+\rho_- +(\rho_+ + \rho_-)a + a^2 = \frac{P^2+Q^2}{4} - \Sigma^2 + 2Ma + a^2,
\ee
where in writing the second equality, we have used the form of $\rho_{\pm}$ as given in \eqn{rho-pm}. Now, since we know from \cite{Lu:2013nt}
that $r_+ r_- = Q^2$ when $P=0$ (also from \cite{Lu:2013nt} that $r_+ r_- = P^2$ when $Q = 0$), so we generalize it to the present case as
\be\label{rpm}
r_+ r_- = P^2 + Q^2,
\ee
which can be used to determine $r_-$ in terms of $r_+$. Equation \eqn{rpm} along with \eqn{rplusrminus} fixes the parameter $a$ as
\be\label{aparameter}
a = -M + \sqrt{M^2 + \Sigma^2 + \frac{3}{4}(P^2+Q^2)},
\ee
where we have used only the plus sign in front of the square root since this reduces to the correct form when $P=0$.
We, thus, find
\be\label{rplus}
r_+ = \rho_+ + a = \sqrt{M^2 + \Sigma^2 - \frac{P^2+Q^2}{4}} + \sqrt{M^2 + \Sigma^2 + \frac{3}{4}(P^2+Q^2)},
\ee
which can be further simplified to give
\be\label{rplussimple}
\frac{1}{4}\left(r_+ - \frac{P^2+Q^2}{r_+}\right)^2 = M^2 + \Sigma^2 - \frac{P^2+Q^2}{4}.
\ee
Note that our intention here is to express $\Sigma$ and $M$ in terms of the horizon radius $r_+$, and for this purpose, we will
use Eq. \eqn{rplussimple}. To eliminate $M^2$ from this equation, we first have from \eqn{dilatoncharge}
  \be \label{p2} M^2 = \frac{\Sigma^2}{3} + \frac{Q^2 - P^2}{8} \frac{\sqrt{3} M \Sigma}{\Sigma^2} - \frac{P^2 + Q^2}{8}.\ee
 We then use \eqn{finalcond} to obtain  $\sqrt{3}M \Sigma$ and substitute it in the above \eqn{p2}
 to obtain $M^2$ in terms of $\Sigma$ and the known charges $P$ and $Q$. Using this expression of $M^2$ in
\eqn{rplussimple} and after some algebraic manipulation, we obtain
\bea \label{eqn1}&& \left(\frac{32}{3}\right)^3 \Sigma^6 - 2 \left(\frac{32}{3}\right)^2 G (r_+) \Sigma^4
  + \frac{32}{3} G^2(r_+) \left(1 + 3 \frac{(Q^2 - P^2)^2}{G^2(r_+)}\right) \Sigma^2 \nn
 && \qquad\qquad\qquad \qquad\qquad \qquad\qquad -
4 (Q^2 - P^2)^2 G(r_+) \left(1 - \frac{Q^2 + P^2}{G(r_+)}\right) = 0,\eea
where we have defined $G(r_+) = 2\left(r_+ - (P^2+Q^2)/r_+\right)^2 + 3(P^2+Q^2)$. Note that \eqn{eqn1} is an equation involving $\Sigma$
and $r_+$, whose explicit solution $\Sigma(r_+)$ is what we want. For this purpose, we further define the following quantities
\be\label{simple}
Y = \frac{32}{3} \frac{\Sigma^2}{G}, \qquad d = \frac{Q^2-P^2}{G} < \frac{1}{3}, \qquad c = \frac{P^2+Q^2}{G} < \frac{1}{3},
\ee
and rewrite  \eqn{eqn1} as
\be\label{eqn2}
Y^3 - 2Y^2 + (1+3d^2)Y - 4d^2(1-c) = 0.
\ee
This is a cubic equation and has three roots in general. We should, of course, take only the real roots. However, as we will see,
even for the real roots, not all of them are allowed. From the definition of $Y$ and for having a well-defined horizon, we conclude
that the allowed solution must be such that
\be\label{Ysoln}
Y = \frac{32}{3}\frac{\Sigma^2}{G} < \frac{32}{3}\frac{\frac{3}{16}(Q-P)^2}{2\left(r_+ - \frac{Q^2+P^2}{r_+}\right)^2 + 3(P^2+Q^2)}
< \frac{2}{3},
\ee
where we have used \eqn{s0} and the definition of $G(r_+)$ as given before. Thus, we conclude that the allowed values of $Y = 32\Sigma^2/(3G)$
must be less than 2/3.

The equation for $Y$, i.e., \eqn{eqn2}, can be solved, and we get the three solutions as follows:
\bea\label{solution1}
Y_1 &=& \frac{2}{3} - \frac{\left( C + \sqrt{C^2 - D}\right)^{1/3} + \left(C - \sqrt{C^2 - D}\right)^{1/3}}{3},\nn
Y_2 &=& \frac{2}{3} + \frac{1}{6}\left[(1 + i \sqrt{3})\left(C + \sqrt{C^2 - D}\right)^{1/3} + (1 - i \sqrt{3})\left(C - \sqrt{C^2 - D}\right)^{1/3}\right],\nn
Y_3 &=& \frac{2}{3} + \frac{1}{6}\left[(1 - i \sqrt{3})\left(C + \sqrt{C^2 - D}\right)^{1/3} + (1 + i \sqrt{3})\left(C - \sqrt{C^2 - D}\right)^{1/3}\right],
\eea
where
\be \label{def2} C = 1 - 27 d^2(1 - 2c), \qquad D = (1 - 9 d^2)^3.\ee
with $c$ and $d$ given in \eqn{simple}. Note that when $C^2 > D$, we have only one real positive root $Y_1$. The other two roots $Y_2$ and $Y_3$
are complex conjugate to each other and must be discarded. Since $Y_1 < 2/3$, it is an allowed solution. On the other hand, when $C^2 < D$, all three
roots are real and positive. In this case, let us define $C = R \cos\theta$ and $\sqrt{D-C^2} = R \sin\theta$, where $R^2=D=(1-9d^2)^3$ and $\cos\theta
=C/\sqrt{D}$ which lies between 0 and 1 and so, $\theta$ lies between 0 and $\pi/2$. With these, \eqn{solution1} can be written as
\bea\label{solution2}
Y_1 &=& \frac{2}{3} - R^{1/3} \frac{e^{i \theta/3} + e^{- i \theta/3}}{3} = \frac{2}{3} \left( 1 - R^{1/3}
\cos\theta/3 \right) > 0,\nn
Y_2 &=& \frac{2}{3} + R^{1/3} \frac{e^{ i (\pi + \theta)/3} + e^{- i (\pi + \theta)/3}} {3} = \frac{2}{3} \left(1 + R^{1/3}
\cos (\pi + \theta)/3\right) > \frac{2}{3},\nn
Y_3 &=&  \frac{2}{3} + R^{1/3} \frac{e^{ - i (\pi - \theta)/3} + e^{ i (\pi - \theta)/3}}{3} = \frac{2}{3} \left(1 + R^{1/3}
\cos (\pi - \theta)/3\right) > \frac{2}{3}.
\eea
Note that since $\theta < \pi/2$, $\cos[(\pi\pm\theta)/3] > 0$, and, therefore, both $Y_2$ and $Y_3$ are greater than 2/3 and, therefore,
should be discarded. However, $Y_1 < 2/3$, and this is the only allowed solution. Thus, we obtain that no matter whether $C^2 > D$
or $C^2 < D$, $Y_1$ is the only allowed solution. We then write
\be\label{sigmarplus}
Y = \frac{32}{3}\frac{\Sigma^2}{G} = \frac{2}{3} - \frac{\left( C + \sqrt{C^2 - D}\right)^{1/3} + \left(C - \sqrt{C^2 - D}\right)^{1/3}}{3},
\ee
where $C$ and $D$ are as given in \eqn{def2}, and $c$ and $d$ in the expression of $C$, $D$ are as given in \eqn{simple}. Also,
$G (r_+) $ is a function of $r_+$ and is given right after \eqn{eqn1}. Equation \eqn{sigmarplus}, therefore, uniquely determines
$\Sigma$ in terms of $r_+$. Further, $M$ can also be expressed in terms of $r_+$ using \eqn{rplussimple} and \eqn{p2} as
\be \label{mp} M = \frac{\Sigma}{\sqrt{3} \, (Q^2 - P^2)} \left( G(r_+) - \frac{32}{3} \Sigma^2\right),\ee
once we have $\Sigma$ in terms of $r_+$. Using \eqn{rplussimple}, \eqn{aparameter}, and the above, we  have
\be \label{aparameternew} a = \frac{\Sigma}{\sqrt{3} \, (Q^2 - P^2)} \left(\frac{32}{3} \Sigma^2 - G(r_+)\right) + \frac{1}{2} \left(r_+ + \frac{Q^2 + P^2}{r_+}\right),
\ee
which will be useful later on.

Once we express $M$ and $\Sigma$ in terms of $r_+$, we can express the entire solution \eqn{bd0d6} in terms of this single parameter
$r_+$ (note that $P$ and $Q$ are fixed charges and, therefore, do not vary). To do this, we replace $\rho$ by $r-a$, where $a$ is given in
\eqn{aparameter}. Then the functions $F(\rho)$, $A(\rho)$, and $B(\rho)$ given in \eqn{fab} can be expressed in terms of $r$ as
\bea\label{fabnew}
F(\rho) &=& \triangle_+ \triangle_- \left(1 - \frac{a}{r}\right)^{-2},\nn
     A (\rho) &=& \left(1 - \frac{r_{A+}}{r}\right)\left(1 - \frac{r_{A-}}{r}\right) \left(1 - \frac{a}{r}\right)^{-2}
\equiv A (r) \left(1 - \frac{a}{r}\right)^{-2},\nn
     B (\rho) &=& \left(1 - \frac{r_{B+}}{r}\right)\left(1 - \frac{r_{B-}}{r}\right) \left(1 - \frac{a}{r}\right)^{-2}
\equiv B (r) \left(1 - \frac{a}{r}\right)^{-2},
\eea
where we have defined, as usual,
\be\label{triangle}
\triangle_\pm = 1 - \frac{r_\pm}{r},
\ee
and
\be r_{A\pm} = a + \rho_{A\pm} = a + \frac{\Sigma}{\sqrt{3}} \pm \sqrt{\frac{P^2 \Sigma/2}{\Sigma -
\sqrt{3} M}},\qquad r_{B\pm} = a + \rho_{B\pm} = a - \frac{\Sigma}{\sqrt{3}} \pm
\sqrt{\frac{Q^2 \Sigma/2}{\Sigma + \sqrt{3} M}}.
\ee
In terms of the new radial coordinate $r$, the configuration \eqn{bd0d6} is
\bea\label{bd0d6inr}
d s^2 &=& \frac{\triangle_+ \triangle_-} {A (r)^{\frac{1}{8}} B (r)^{\frac{7}{8}}} d t^2 + \left(\frac{B(r)}{A (r)}\right)^{\frac{1}{8}}
\Sigma_{i = 1}^6 d x_i^2 + A^{\frac{7}{8}} (r) B^{\frac{1}{8}} (r) \left(\frac{d r^2}{\triangle_+ \triangle_-} + r^2 d\Omega_2^2\right),\nn
A_{[1]} &=& i e^{- 3 \phi_0/4} Q \left[\frac{1 - \frac{ \sqrt{3} a  +  \Sigma}{\sqrt{3} r_+}}{r_+ B (r_+)} - \frac{1 - \frac{\sqrt{3} a
+ \Sigma}{\sqrt{3} r}}{r B (r)}\right] d t,\nn
A_{[7]} &=& i e^{ 3 \phi_0/4} P \left[\frac{1 - \frac{ \sqrt{3} a - \Sigma}{\sqrt{3} r_+}}{r_+ A (r_+)} - \frac{1 - \frac{\sqrt{3} a -
\Sigma}{\sqrt{3} r}}{r A (r)}\right] dt \wedge d x^1 \wedge \cdots \wedge d x^7,\nn
e^{2 (\phi - \phi_0)} &=& \left(\frac{B (r)}{A (r)}\right)^{3/2}.
\eea
Assuming that this configuration has a well-defined horizon at $r=r_+$, the metric can be made free of conical singularity at the
horizon if the Euclidean time ``$t$'' is compact with periodicity
\be\label{period}
\beta^{\ast} = \frac{4 \pi r_+^2 \sqrt{A(r_+) B(r_+)}}{r_+ - r_-}.
\ee
This is the inverse of the temperature of the black $D6/D0$ system at infinity. The inverse of the local temperature at a given $r$, which is
important for the analysis of the phase structure, is given as
\be\label{periodlocal}
\beta (r) = \sqrt{\frac{A(r_+) B(r_+)}{A^{1/8}(r) B^{7/8}(r)}} \frac{4\pi r_+ (\triangle_+ \triangle_-)^{1/2}}{1 - \frac{r_-}{r_+}}.
\ee
As mentioned in \cite{Lu:2010xt, Lu:2013nt},  we should use physical radius $\bar{r} = \sqrt{A^{7/8}(r) B^{1/8}(r)}\,r$ instead of
the coordinate radius $r$ and also the physical paramaters $\bar{r}_{\pm} =  \sqrt{A^{7/8}(r) B^{1/8}(r)}\, r_{\pm}$ at a given $r$. Note that with these,
$\triangle_{\pm}(r) = \triangle_{\pm}(\bar{r})$. For other related parameters, their physical correspondences should also be used accordingly. For example,
given $r_+ r_- = Q^2 + P^2$ from \eqn{rpm}, the physical $\bar Q = \sqrt{A^{7/8}(r) B^{1/8}(r)}\, Q$ and so it is for $\bar P$.   Now, in terms of the physical
coordinate, the inverse of the local temperature \eqn{periodlocal} at the given
radius $\bar r$
takes the form
\be\label{periodlocal1}
\beta (\bar r) = \sqrt{\frac{A(\bar{r}_+) B(\bar{r}_+)}{A(\bar{r}) B(\bar r)}} \frac{4\pi \bar{r}_+ (\triangle_+(\bar r) \triangle_-(\bar r))^{1/2}}{1 -\frac{
\bar{r}_-}{\bar{r}_+}}.
\ee

To study the equilibrium thermodynamics \cite{Gibbons:1976ue} in the canonical ensemble, as mentioned in the beginning of this section,
the allowed configuration must be placed in a cavity with fixed radius $\bar r = \bar{r}_B > \bar{r}_+$. The other quantities which are held
fixed are the cavity temperature, $1/\bar \beta$, the
physical periodicity of each $x^i$, for $i=1,2,\ldots,6$, the dilaton value $\bar{\phi}$ on the surface of the cavity (at $\bar{r} = \bar{r}_B$),
and the charges enclosed in the cavity $\bar{P}$, $\bar{Q}$. In equilibrium, these values are taken to be equal to the corresponding values of the
allowed configuration enclosed in the cavity. Note that the usual asymptotic value of dilaton $\phi_0$ is not fixed but is expressed in terms of the
fixed $\bar \phi$ via \eqn{bd0d6inr} as $e^{\phi_0} = e^{\bar\phi} (A (\bar r_B)/B(\bar r_B))^{3/4}$ where we have set $\phi(\bar r_B) = \bar\phi$. In what
follows, we use a ``bar'' above the symbol to denote the corresponding physical/or fixed parameter.

In the canonical ensemble, the stability analysis can be performed using the Helmholtz free energy $F$ of the system under consideration which, to leading order, is given as $F = I_E/\bar \beta$ with  $I_E$ the Euclidean action \cite{Gibbons:1976ue}. We actually ask this question: in the given condition set by the canonical ensemble, i.e., with fixed $\bar r_B, \bar Q, \bar P, \bar \phi, \bar \beta$, what thermodynamically stable phase of charged black $D6/D0$ in the cavity can exist? Note that in the canonical ensemble, the only variable for this system is the horizon size $\bar r_+$, and so the local minimum of $F$ with respect to $\bar r_+$ will determine the local stability of the underlying system. With $\bar \beta$ fixed, this can, in turn, be determined from the local minimum of $I_E$ with respect to $\bar r_+$.

Following our previous work \cite{Lu:2010xt,Lu:2012rm,Lu:2013nt}, the Euclidean action $I_E$ for the charged black $D6/D0$ configuration \eqn{bd0d6inr}  in the canonical ensemble as specified above  can be explicitly computed and its so-called reduced Euclidean action $\tilde I_E$ is actually relevant for the above-mentioned stability analysis and is given as
\bea \label{raction} \tilde I_E &\equiv& \frac{2 \kappa^2 I_E}{4\pi \Omega_2 \bar V_6 \bar r_B^2}\nn
&=& \bar b  \left[\frac{q_+^2 - q_-^2}{16 \left(\triangle_+ \triangle_-\right)^{1/2}} \left( \frac{A_- (\bar r_+)}{x A(\bar r_+)} - \frac{A_- (\bar r_B)}{A (\bar r_B)}\right) + \frac{7 (q_+^2 + q_-^2)}{16 \left(\triangle_+ \triangle_-\right)^{1/2}}\left(\frac{A_+ (\bar r_+)}{x B(\bar r_+)} - \frac{A_+ (\bar r_B)}{B (\bar r_B)}\right)\right.\nn
&\,& \quad + \left. 4 - 2 \left(\frac{\triangle_+}{\triangle_-}\right)^{1/2} - \frac{(\triangle_+ \triangle_-)^{1/2}}{4}\left(\frac{7 A_+ (\bar r_B)}{A (\bar r_B)} + \frac{A_- (\bar r_B)}{B (\bar r_B)}\right)\right] - x^2 \sqrt{\frac{A(\bar r_+) B (\bar r_+)}{A (\bar r_B) B (\bar r_B)}},\nn
\eea
where $\Omega_n$ denotes the volume of a unit $n$-sphere, the physical volume $\bar V_6$ is related to the coordinate volume $V_6^* \equiv \int d x^1 d x^2 \cdots d x^6$ via $\bar V_6 = (B(\bar r_B)/A(\bar r_B))^{3/8} V_6^*$ from the metric given in \eqn{bd0d6inr}, and $\kappa$
is a constant with $1/ (2 \kappa^2)$ appearing in front of the Hilbert-Einstein action in canonical frame but containing no asymptotic string coupling $g_s = e^{\phi_0}$. Also in the above, as usual, for simplicity we introduce the so-called reduced quantities
at the fixed radius $\bar r = \bar{r}_B$ by the relations,
\be\label{reduced}
x \equiv \frac{\bar{r}_+}{\bar{r}_B} < 1, \qquad \bar b \equiv \frac{\bar \beta}{4\pi\bar{r}_B}, \qquad q_+ \equiv \frac{\bar Q_+}{\bar r_B} < x, \qquad  q_- \equiv \frac{\bar Q_-}{\bar r_B} < q_+,
\ee
with $\bar Q_+^2 = \bar Q^2 + \bar P^2, \quad \bar Q_-^2 = \bar Q^2 - \bar P^2$ (assuming $Q > P$).\footnote{As mentioned earlier, we assume $Q > P$ in our discussion, and $Q < P$ case can be obtained from the $Q > P$ by the duality following the discussion after \eqn{dilatoncharge}.} In \eqn{raction}, we also define
\be \label{Apm} A_\pm (\bar r) = 1 - \frac{\sqrt{3}\bar a \pm \bar \Sigma}{\sqrt{3} \bar r}.\ee
In terms of these reduced quantities, the functions $A(\bar r)$, $A(\bar{r}_+)$, $B(\bar r)$, and $B(\bar{r}_+)$ can be written as
\bea \label{abdefd}
A (\bar r_B) &=& A^2_+ (\bar r_B) + \frac{q_-^2}{4} -\frac{2}{3}\left(\frac{\bar \Sigma}{\bar r_B}\right)^2  + \frac{2}{3 q_-^2}\left[\frac{32}{3} \left(\frac{\bar\Sigma}{\bar r_B}\right)^2 - g (x) \right]\left(\frac{\bar\Sigma}{\bar r_B}\right)^2,\nn
x^2 A (\bar r_+) &=& \left(x - \frac{\sqrt{3} \bar a + \bar\Sigma}{\sqrt{3} \bar r_B}\right)^2 + \frac{q_-^2}{4} -\frac{2}{3}
\left(\frac{\bar\Sigma}{\bar r_B}\right)^2 + \frac{2}{3 q_-^2}\left[\frac{32}{3} \left(\frac{\bar\Sigma}{\bar r_B}\right)^2 - g(x) \right]\left(\frac{\bar\Sigma}{\bar r_B}\right)^2,\nn
B(\bar r_B) &=& A^2_- - \frac{q_-^2}{4} -\frac{2}{3} \left(\frac{\bar\Sigma}{\bar r_B}\right)^2 - \frac{2}{3 q_-^2}\left[\frac{32}{3} \left(\frac{\bar\Sigma}{\bar r_B}\right)^2 -  g(x) \right]\left(\frac{\bar\Sigma}{\bar r_B}\right)^2,\nn
x^2 B (\bar r_+) &=& \left(x - \frac{\sqrt{3} \bar a - \bar\Sigma}{\sqrt{3} \bar r_B}\right)^2 - \frac{q_-^2}{4} -\frac{2}{3}
\left(\frac{\bar\Sigma}{\bar r_B}\right)^2 - \frac{2}{3 q_-^2}\left[\frac{32}{3} \left(\frac{\bar\Sigma}{\bar r_B}\right)^2 - g(x) \right]\left(\frac{\bar\Sigma}{\bar r_B}\right)^2,\nn
\eea
where $g(x)=2x^2\left(1-\frac{q_+^2}{x^2}\right)^2 + 3 q_+^2$.

Note that $I_E  = \bar\beta\, E - S$ where $E$ is the internal energy of the system, and $S$ is the entropy. In terms of the reduced Euclidean action and the reduced quantities, we have $\tilde I_E (x; q_+, q_-) = \bar b\, \tilde E_{q_+,\, q_-} (x) - \tilde S_{q_+,\, q_-} (x)$. By comparing this with \eqn{raction}, one can read both $\tilde E_{q_+,\, q_-} (x)$ and $\tilde S_{q_+,\, q_-} (x)$, explicitly. As stressed earlier, in the canonical ensemble, both $q_+, \, q_-$ are fixed; the only variable is the reduced horizon size $x$, and so we have
\be \label{extrem} \frac{d \tilde I_E}{d x} = \frac{d \tilde E_{q_+,\, q_-} (x)}{d x\qquad} \left(\bar b - b_{q_+,\, q_-} (x)\right),\ee
where \be
\label{bf} b_{q_+,\,q_-} (x) \equiv \frac{d S_{q_+,\,q_-} (x)/d x}{d E_{q_+,\, q_-} (x)/d x}.\ee
From \eqn{extrem}, we have
\be \frac{d \tilde I_E}{d x} = 0 \Rightarrow  b_{q_+,\, q_-} (\bar x) = \bar b,\ee
where the extremal condition of $\tilde I_E$ is nothing but the thermal equilibrium of the charged black system, with a horizon size $x =\bar x$ determined by the above equation, with the cavity with a preset reduced temperature $1/\bar b$. At $x = \bar x$,  we further have
\be \left.\frac{d^2 \tilde I_E}{d x^2}\right|_{x = \bar x} = - \left.\frac{d \tilde E_{q_+,\, q_-} (x)}{d x\qquad}\right|_{x = \bar x}  \left.\frac{ d b_{q_+,\, q_-} (x)}{d x}\right|_{x = \bar x}.\ee Since $E_{q_+, \, q_-} (x)$ is an increasing function of $x$ for $0< x < 1$,  the minimum of $\tilde I_E$ implies then, as usual, the negative slope of $b_{q_+,\, q_-} (x)$ at $x = \bar x$. So, the function $b_{q_+,\, q_-} (x)$ is the key for determining the underlying phase structure. The explicit expression of $b_{q_+,\, q_-} (x)$ can be obtained as described above but with a lengthy computation, and it turns out, as expected, to be nothing but the $\beta (\bar r)$ given in \eqn{periodlocal1} at $\bar r = \bar r_B$ and expressed in terms of the reduced quantities. It is given as\footnote{Note that with $\bar r_B, q_+,
\, q_-$ fixed, $\beta (\bar r)$ in \eqn{periodlocal1} at $\bar r = \bar r_B$ is the only function of the reduced horizon size $x$ and $b_{q_+,\, q_-} (x) \equiv \frac{\beta (\bar r_B)}{4 \pi \bar r_B}$.}
\be\label{bfdef} b_{q_+,\, q_-} (x) = \sqrt{\frac{A (\bar r_+) B (\bar r_+)}{A(\bar r_B) B (\bar r_B)}}\,\,x \left(1 - x\right)^{1/2} \left(1 -
\frac{q_+^2}{x}\right)^{1/2} \left(1 - \frac{q_+^2}{x^2}\right)^{-1},\ee
where functions $A(\bar r_+), B(\bar r_+)$, and $A (\bar r_B), B(\bar r_B)$ are given in \eqn{abdefd}.
From our experience \cite{Lu:2010xt,Lu:2012rm,Lu:2013nt}, we know that the existence of the universal van der Waals$-$Maxwell liquid-gas$-$type phase structure depends crucially on whether the  $b_{q_+, \, q_-} (x)$ blows up at $x \to q_+$, i.e., the extremal limit.

For this we need to examine the behaviors of $A(\bar r_B)$, $A(\bar{r}_+)$, $B(\bar r_B)$, and $B(\bar{r}_+)$. When $q_+ = q_-$, i.e.,
$P=0$, we can obtain from \eqn{sigmarplus}, $\bar\Sigma/\bar r_B = \sqrt{3}q_+/(4x)$ and from there we obtain $\bar a/\bar r_B =
3q_+^2/(4x)$. We then have from \eqn{abdefd},
\be A (\bar r_B) = \left(1 - \frac{q_+^2}{x}\right)^2,\quad B(\bar r_B) = 1 - \frac{q_+^2}{x},\quad A(\bar r_+) = \left(1 -
\frac{q_+^2}{x^2}\right)^2, \quad B(\bar r_+) = 1 - \frac{q_+^2}{x^2}.\ee
Substituting these into \eqn{bfdef}, we get
\be\label{bfdefspl}
b_{q_+,\,q_+} (x) = x (1-x)^{1/2} \left(1-\frac{q_+^2}{x}\right)^{-1} \left(1-\frac{q_+^2}{x^2}\right)^{1/2}.
\ee
This is precisely the result obtained in \cite{Lu:2010xt} for charged black $D6$ branes  when $D=10$. Note here that the structure of
the inverse of the reduced local temperature
\eqn{bfdefspl} for $D6$ branes is different (it is actually regular as $x \to q_+$) from the structure obtained for $D6/D0$ system (it
blows up in the extremal limit) for a special case with $Q=P$ in
\cite{Lu:2013nt}. We now look at the case when $q_- < q_+$. For this, let us find the expressions
for $\bar\Sigma/\bar r_B$ and $\bar a/\bar r_B$ first. From the solution of $Y$ in \eqn{sigmarplus}, we find
\be \label{sdef}\frac{\bar \Sigma}{\bar r_B} = \frac{1}{4} g^{\frac{1}{2}}(x)
\left[1 - \frac{\left( C + \sqrt{C^2 - D}
\right)^{\frac{1}{3}} + \left(C - \sqrt{C^2 - D}\right)^{\frac{1}{3}}}{2}\right]^{\frac{1}{2}}.
\ee
From \eqn{aparameternew}, we have
\be \label{adef}
\frac{\bar a}{\bar r_B} = \frac{1}{2} x \left(1 + \frac{q_+^2}{x^2}\right) + \frac{\sqrt{3}}{q_-^2}\left[\frac{32}{9}
\frac{\bar \Sigma^2}{\bar r_B^2} - \frac{1}{3} g (x)\right]\frac{\bar \Sigma}{\bar r_B}.
\ee
Note that the parameters $c$ and $d$ given in \eqn{simple} can now be written as $c(x) = q_+^2/g(x)$ and $d(x) = q_-^2/g(x)$, and,
therefore, as $x \to q_+$, $g(x) \to 3q_+^2$ and so $c(x)$, $d(x)$, as well as $C(x)$, $D(x)$ [given in \eqn{def2}] go to
\bea\label{cdCD}
c(x) & \to & \frac{1}{3}, \qquad\qquad d(x) \to \frac{q_-^2}{3q_+^2},\nn
C(x) & \to & 1 - \frac{ q_-^4}{q_+^4}, \quad D(x) \to \left(1 - \frac{q_-^4}{q_+^4}\right)^3.
\eea
Now, substituting these in \eqn{sdef} and in \eqn{adef}, we find that $\bar\Sigma/\bar r_B \approx q_+ (q_-/q_+)^2 /6$ and $\bar a/\bar r_B \approx q_+/2$ both are regular as
$x \to q_+$. Using \eqn{abdefd}, we have then
\be A (\bar r) \approx \left(1 - \frac{q_+}{2}\right)^2 + \frac{q_-^2}{12},\quad B (\bar r) \approx \left(1 - \frac{q_+}{2}\right)^2 - \frac{q_-^2}{12} > 0,\quad A(\bar r_+) = B(\bar r_+) \approx \frac{1}{4},\ee which are all regular as $x \rightarrow q_+$.  From these, we have
\be \sqrt{\frac{A (\bar r_+) B (\bar r_+)}{A (\bar r) B (\bar r)}} \approx \frac{1}{4 \left[\left(1 - \frac{q_+}{2}\right)^4 - \frac{q_-^4}{12^2}\right]^{1/2}},\ee which is also regular. Thus, the singular structure of $b_{q_+,\, q_-} (x)$ given
in \eqn{bfdef} for $q_+ > q_-$ as $x \rightarrow q_+$ is the same as the $Q=P$ case studied previously in \cite{Lu:2013nt}.\footnote{Note that for the $Q=P$ case,
$A(\bar r_B) = B(\bar r_B)= A(\bar{r}_ +)= B(\bar{r}_+) = 1$, and so, in that case, the inverse of the reduced temperature has the form given in
\eqn{bfdef} without the first square-root factor.} Therefore, the phase structure essentially remains
the same as in the $Q=P$ case; however, we expect the phase structure to be much richer here [since the first square-root factor in \eqn{bfdef} will change the details of the phase structure], similar to that of the $D5/D1$ system. Given the complicated dependence of $b_{q_+,\, q_-} (x)$ on $x$ (also on $q_+, q_-$ as well) as given
in \eqn{bfdef} with $A (\bar r_B), B(\bar r_B), A(\bar{r}_ +)$, and  $B(\bar{r}_+)$  given in \eqn{abdefd}, unlike the $D5/D1$ system, we are unable to give an analytic analysis of the underlying phase structure, in particular, the critical phenomenon, here. However, we can still say something about the critical charge $(q_{+ c}, q_{-c})$ in the present case vs the $q_c  = \sqrt{5} - 2 \approx 0.24$ in the case of $Q = P$ (or $q_- = 0$) given in \cite{Lu:2013nt}.
For each $q_- \neq 0$ with $ 0 < q_- < q_+$, we expect the corresponding critical charge
$q_{+ c} > q_c = \sqrt{5} - 2$ for the following reason. For this, let us denote the first square-root factor in \eqn{bfdef} as $ w_{q_+,\,q_-} (x)$ and the remaining as $b_{q_+} (x)$. We can then rewrite
\be \label{bfnew} b_{q_+,\, q_-} (x) = w_{q_+,\, q_-} (x) b_{q_+} (x).\ee
Note that for $q_- = 0$, the corresponding inverse of the reduced local temperature is precisely the same as the $b_{q_+} (x)$ since now $w_{q_+,\, q_-} (x) = 1$  \cite{Lu:2013nt}.
We also know that for $q_- \neq 0$
$w_{q_+,\, q_-} (x\rightarrow 1) \rightarrow 1$ from \eqn{abdefd} and $w_{q_+,\, q_-} (x\rightarrow q_+) < 1$. Actually, $w_{q_+,\, q_-} (x)$ is an increase function of $x$ for $q_+ < x < 1$. We use two figures with different pairs of ($q_+, q_-$) values for showing this.
\begin{figure}[htp]
\begin{center}
\includegraphics[scale=0.40, angle= -90]{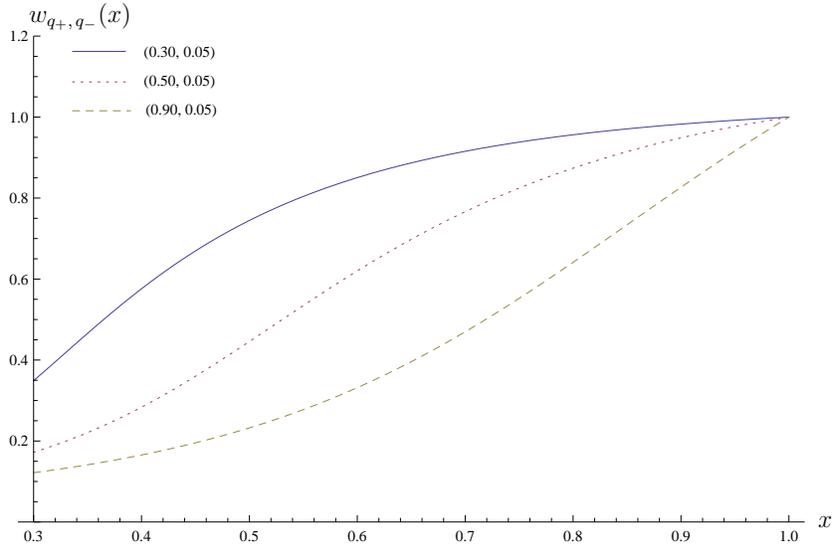}
\end{center}
\caption{The behavior of $w_{q_+,\, q_-} (x)$ vs $x$ for a given $q_- = 0.05$ and three different $q_+ = 0.30, 0.50, 0.90$, respectively.}
\end{figure}
\begin{figure}[htp]
\begin{center}
\includegraphics[scale=0.40, angle= -90]{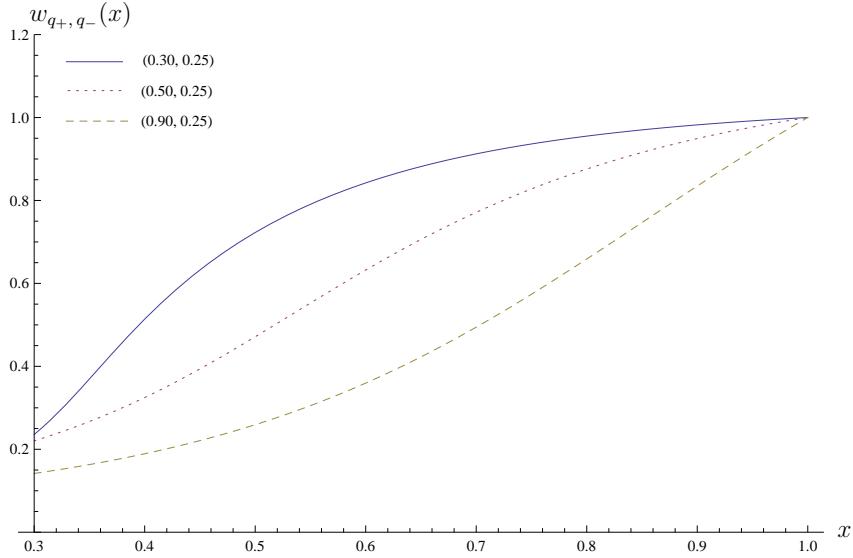}
\end{center}
\caption{The behavior of $w_{q_+,\, q_-} (x)$ vs $x$ for a given $q_- = 0.25$ and three different $q_+ = 0.30, 0.50, 0.90$, respectively.}
\end{figure}
From Figs.1 and 2, we see that $w_{q_+,\, q_-} (x \rightarrow 1) \rightarrow 1$, and close to $x \rightarrow 1$, this function is more sensitive to  $q_+$ values, while close to $x\rightarrow q_+$, it is sensitive to both $q_+$ and $q_-$ values. For the $q_- = 0$ case, $b_{q_+} (x)$ gives the corresponding  critical charge $q_c = \sqrt{5} - 2$ which is determined by requiring that both its first and second derivatives vanish \cite{Lu:2013nt}. For this critical $q_c$, we also have a critical reduced horizon size  $ x_c = 5 - 2 \sqrt{5}$ \cite{Lu:2013nt}, and if $x$ is close to $x_c$, we have $b_{q_c} (x > x_c) = b_{q_c} (x < x_c)$ up to the order  $\cal{O}$ ( $(x - x_c)^3$). Now, for $q_+ = q_c$ and $q_- \neq 0$, we must have $b_{q_+,\, q_-} (x < x_c) = w_{q_+,\, q_-} (x < x_c) b_{q_+} (x < x_c) < b_{q_+,\, q_-} (x > x_c) = w_{q_+,\, q_-} (x > x_c) b_{q_+} (x > q_c)$ since $w_{q_+,\,q_-} (x < x_c) < w_{q_+,\, q_-} (x > x_c)$ even though we still have $b_{q_+} (x > x_c) \approx b_{q_+} (x < x_c)$ in the sense described above. Given our experience about the van der Waals$-$Maxwell liquid-gas$-$type phase structure\cite{Lu:2010xt, Lu:2013nt}, the $q_+ = q_c = \sqrt{5} - 2$ is less than the actual critical charge $q_{+c}$ for the present system since, otherwise, we should have $b_{q_+, \, q_-} (x < x_c) \ge b_{q_+,\, q_-} (x > x_c)$. In other words, the critical charge $q_{+c} > q_c = \sqrt{5} - 2 \approx 0.24$ in the case of $q_{-c} \neq 0$. In the following, we give a few figures to show this and also indicate how the underlying phase structure depends on both $q_+$ and $q_-$.
\begin{figure}[htp]
\begin{center}
\includegraphics[scale=0.40, angle= -90]{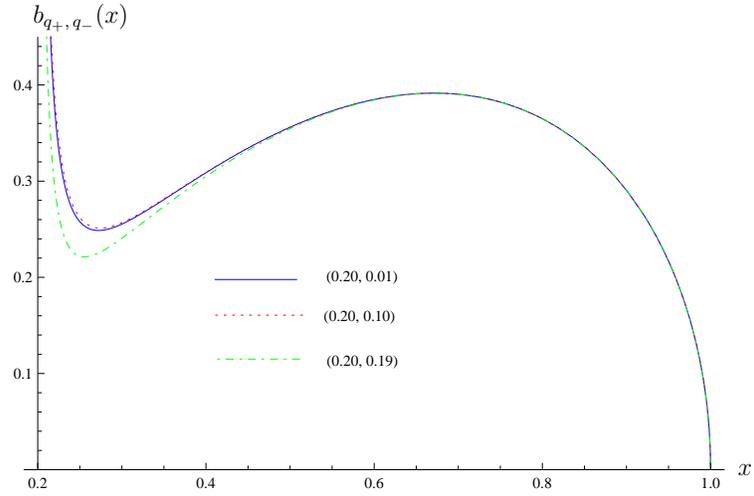}
\caption{The behavior of $b_{q_+,\, q_-} (x)$ vs $x$ for a given $q_+ = 0.20$ and three different $q_- = 0.01, 0.10, 0.19$, respectively.}
\end{center}
\end{figure}
\begin{figure}[htp]
\begin{center}
\includegraphics[scale=0.40, angle= -90]{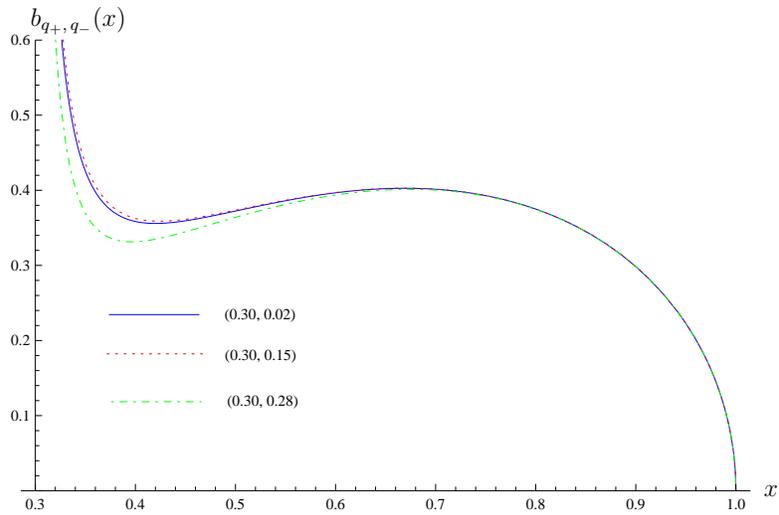}
\end{center}
\caption{The behavior of $b_{q_+,\, q_-} (x)$ vs $x$ for a given $q_+ = 0.30$ and three different $q_- = 0.02, 0.15, 0.28$, respectively.}
\end{figure}

Figures 3$-$5 each consider the behavior of $b_{q_+,\, q_-} (x)$ vs $x$ for a given $q_+$ value and three different $q_-$ values. Once again, we see that in each case, $q_-$ has its influence on $b_{q_+,\, q_-} (x)$ mainly for $x$ close to the end of $x = q_+$ while it has almost no influence for $x$ close to the other end of $x = 1$. In each case, we consider a small $q_-$ corresponding to $Q \gtrsim P$, a characteristic value $q_-$ corresponding to $Q > P$, and a $q_- \lesssim q_+$ corresponding to $P \gtrsim 0$.
\begin{figure}[htp]
\begin{center}
\includegraphics[scale=0.40, angle= -90]{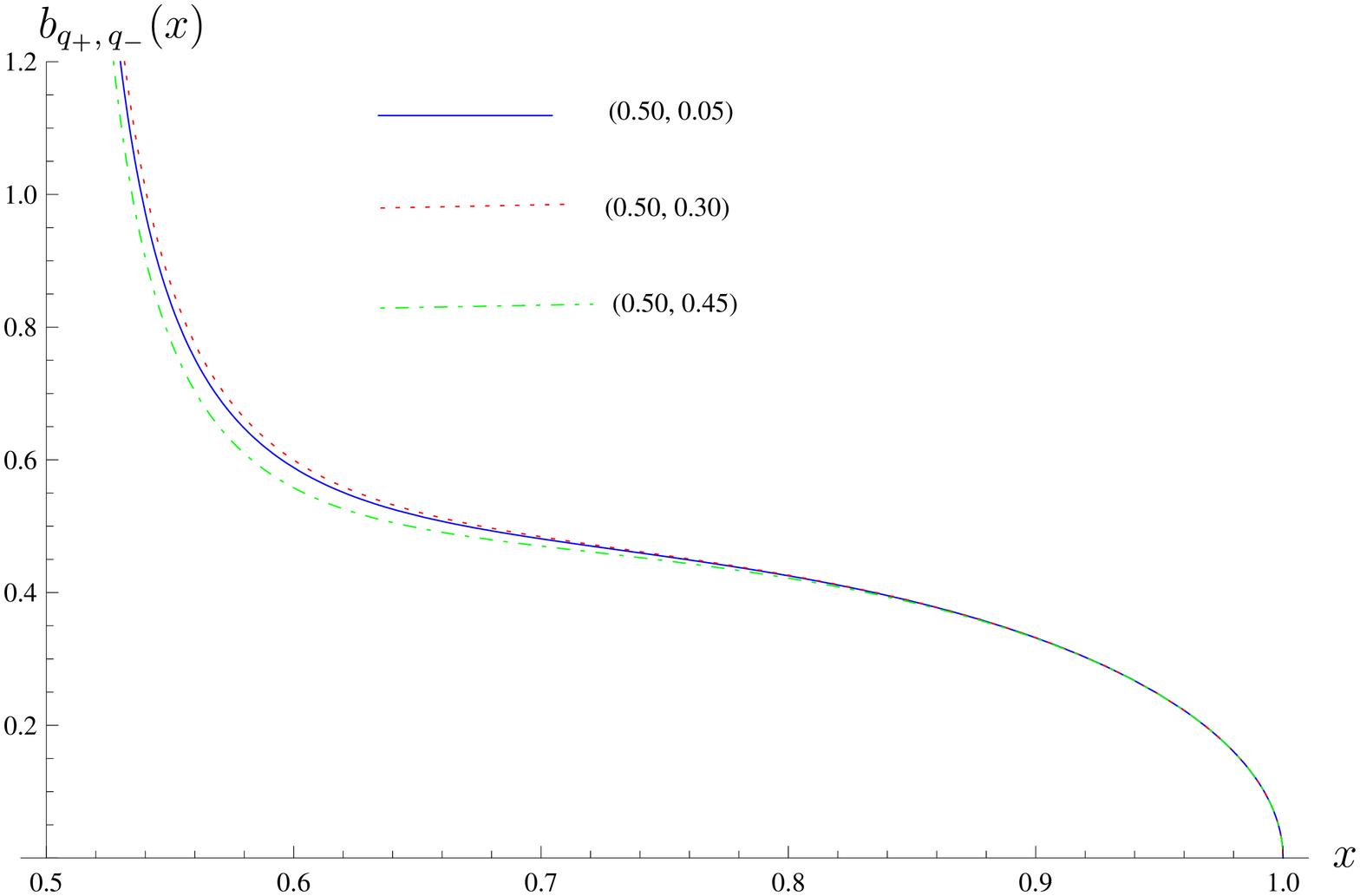}
\end{center}
\caption{The behavior of $b_{q_+,\, q_-} (x)$ vs $x$ for a given $q_+ = 0.50$ and three different $q_- = 0.05, 0.30, 0.45$, respectively.}
\end{figure}

 \begin{figure}[htp]
\begin{center}
\includegraphics[scale=0.40, angle= -90]{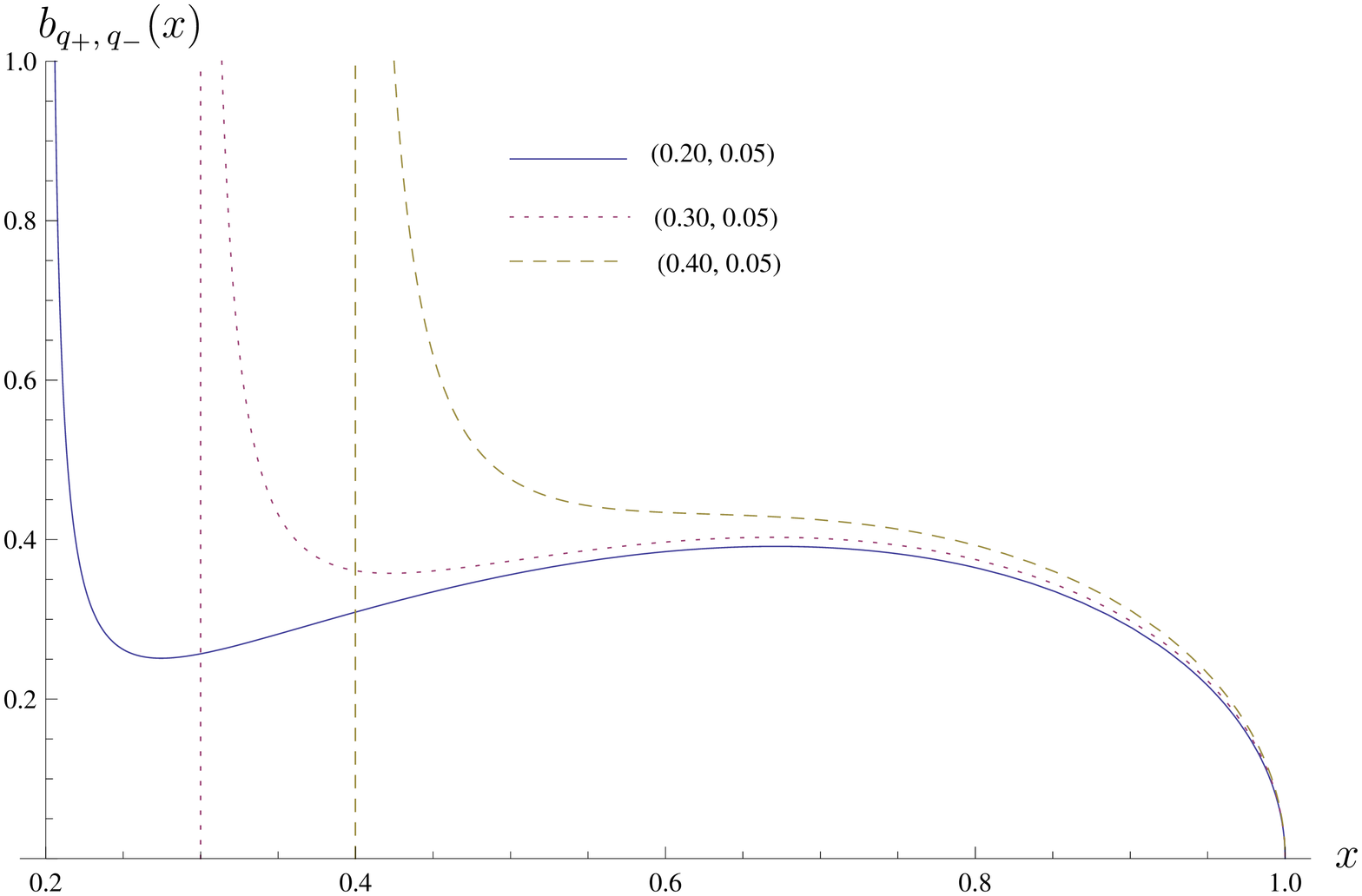}
\end{center}
\caption{The behavior of $b_{q_+,\, q_-} (x)$ vs $x$ for a given $q_- = 0.05$ and three different $q_+ = 0.20, 0.30, 0.40$, respectively.}
\end{figure}
\begin{figure}[htp]
\begin{center}
\includegraphics[scale=0.40, angle= -90]{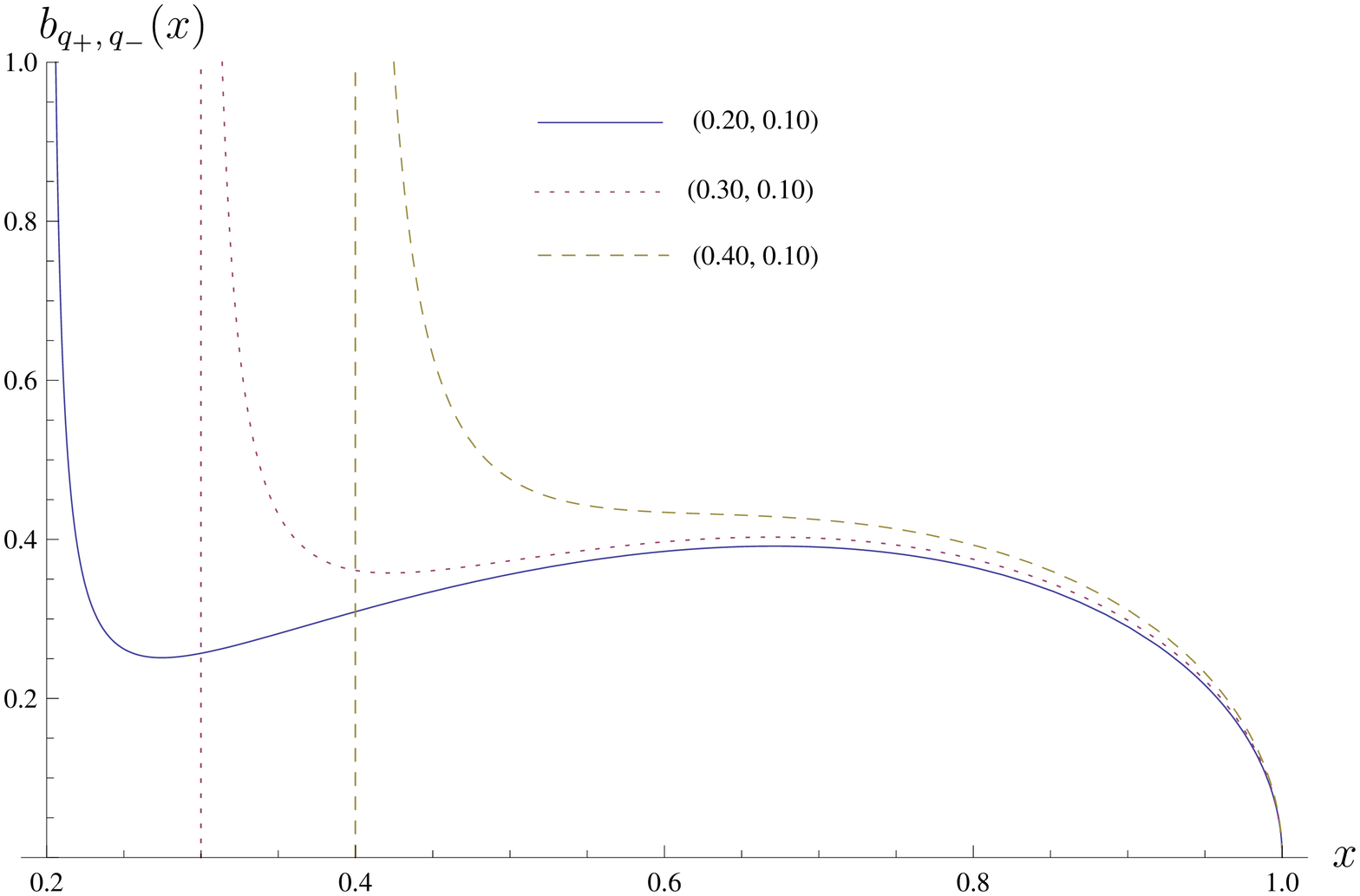}
\end{center}
\caption{The behavior of $b_{q_+,\, q_-} (x)$ vs $x$ for a given $q_- = 0.10$ and three different $q_+ = 0.20, 0.30, 0.40$, respectively.}
\end{figure}
 \begin{figure}[htp]
\begin{center}
\includegraphics[scale=0.40, angle= -90]{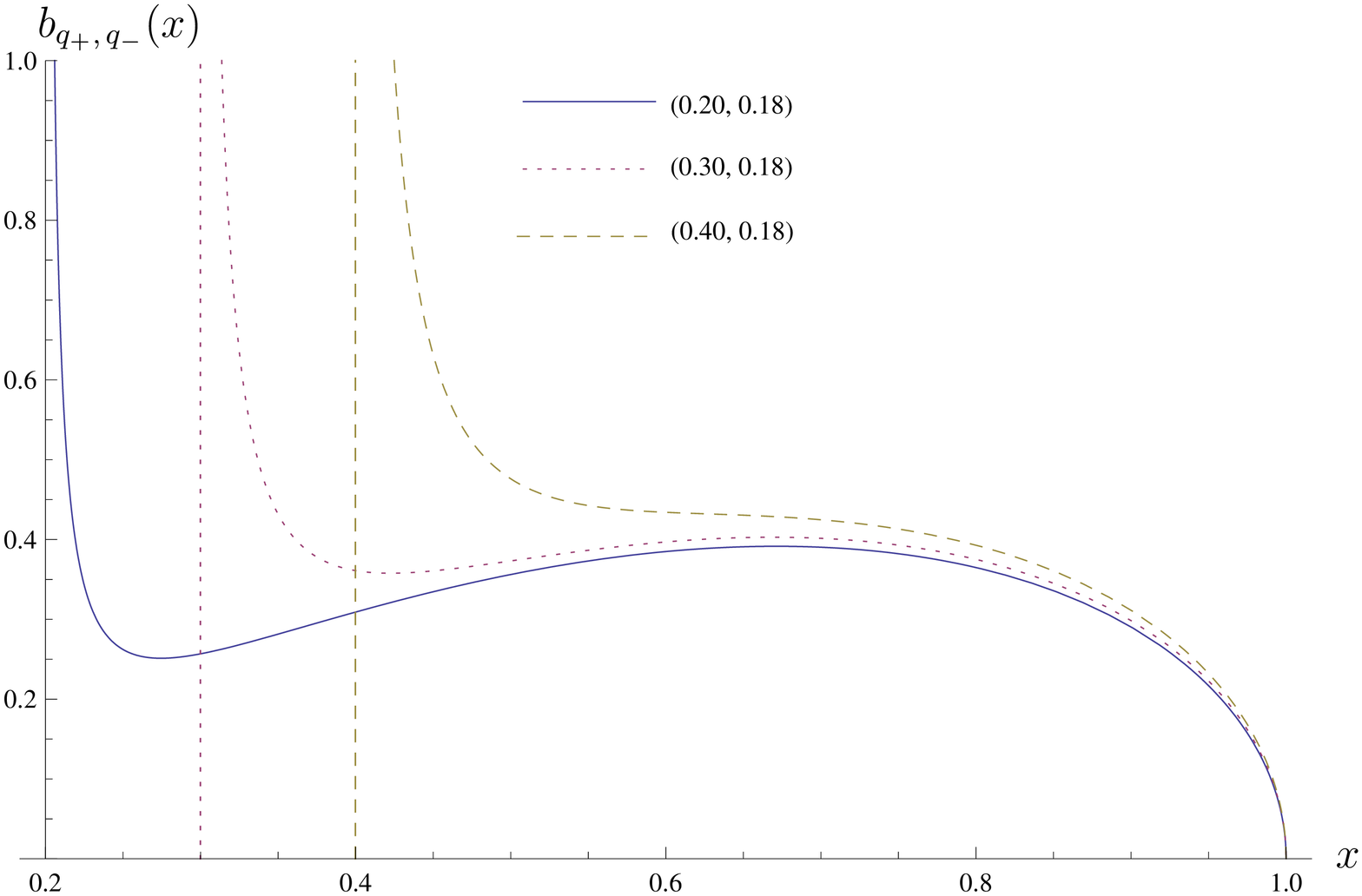}
\end{center}
\caption{The behavior of $b_{q_+,\, q_-} (x)$ vs $x$ for a given $q_- = 0.18$ and three different $q_+ = 0.20, 0.30, 0.40$, respectively.}
\end{figure}
\begin{figure}[htp]
\begin{center}
\includegraphics[scale=0.40, angle= -90]{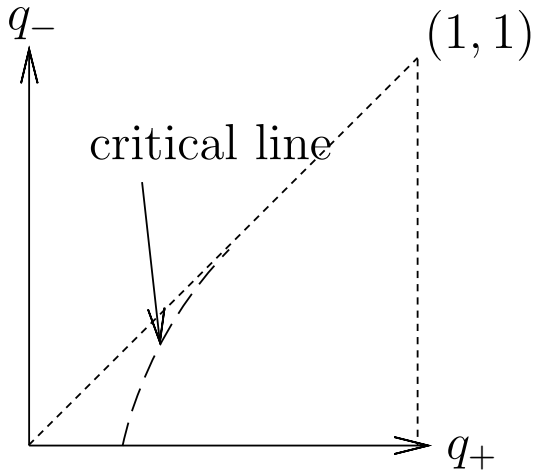}
\end{center}
\caption{The two-dimensional region of allowed reduced charge $q_+$ and $q_-$.}
\end{figure}
   From Figs. 4 and 5, we see that the corresponding critical charge $q_{+c}$ falls between $0.30$ and  $0.50$, consistent with what we discuss above about $q_{+c} > q_c =\sqrt{5} - 2$. Figure 5 indicates that the influence of $q_-$ on the behavior of $b_{q_+,\, q_-} (x)$ becomes less important even for $x$ close to the end of $x = q_+$ when $q_+ > q_{+c}$.  Here we give three more figures (Figs. 6$-$8), each of which is now for a given $q_-$ value and three different $q_+$ values, to indicate what has been said about the critical charge $q_{+c}$.
  Similar to
  the $D5/D1$ system \cite{Lu:2012rm}, the charges $q_+$ and $q_-$ span a two-dimensional region bounded by $q_+ = q_-; q_- = 0, 0 \le  q_+ \le 1$ and $q_+ = 1, 0\le q_- \le  1$, as shown in Fig. 9. In this figure, we draw also a characteristic critical line determined by the vanishing of the first and the second derivatives of $b_{q_+,\, q_-} (x)$ with respect to $x$. As mentioned above, the complicated expression of $b_{q_+,\, q_-} (x)$ makes it impossible for us to give an analytic analysis of this critical line, unlike the case of $D5/D1$ system \cite{Lu:2012rm}.  As discussed above already, this critical line starts at $q_{+c} = q_c = \sqrt{5} - 2, q_{-c} = 0$,  and once $q_{-c} > 0$, $q_{+c} > q_c = \sqrt{5} - 2$, but the ending point cannot be determined analytically since $q_+ = q_-$ can never be even a fake ``critical point''\footnote{In the sense the first and second derivatives of $b_{q_+,\, q_-} (x)$ vanish as for the $D5/D1$ system even though this point is not a true critical point.}  since this corresponds to the $P = 0$ case, and the corresponding system has no van der Waals$-$Maxwell liquid-gas$-$type phase structure\cite{Lu:2013nt}. Our numerical tries indicate that the ending point is around $q_{+} = 0.52$ with a $q_-$ very close to this value but not reaching the $q_- = q_+$ line. Figure 10 gives a flavor of this for $(q_{+c},\, q_{-c}) = (0.520000000,  0.519999999)$. From this, one can see that the critical size $x_c$ should fall between $0.52$ and $0.64$. This critical line separates the $(q_+, q_-)$ region into two parts, the small one on the left and the large one on the right, as shown in Fig. 9.
  \begin{figure}[htp]
\begin{center}
\includegraphics[scale=0.40, angle= -90]{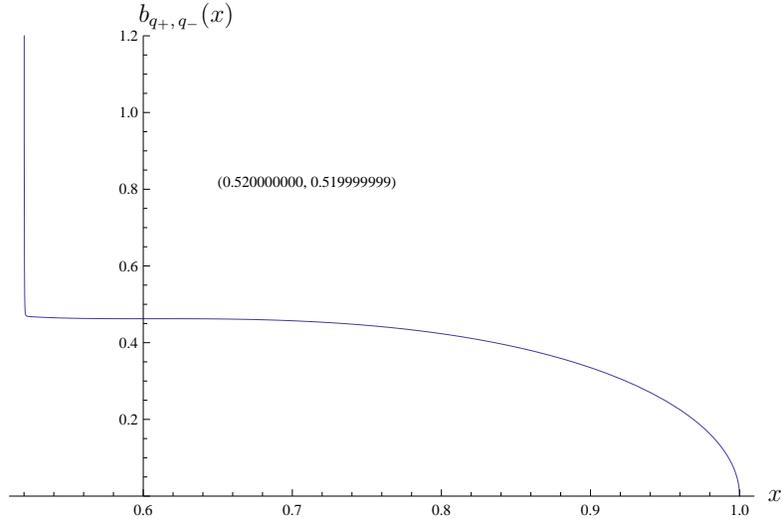}
\end{center}
\caption{The critical behavior of $b_{q_+,\, q_-} (x)$ vs $x$ for a given pair of $(q_+,\, q_-) = (0.520000000,  0.519999999)$.}
\end{figure}

  For each given pair of $(q_+, q_-)$ with $q_- < q_+$ in the left part,  $b_{q_+, \, q_-} (x)$ has a minimum $b_{\rm min}$ and a maximum $b_{\rm max}$ in the region of $q_+ < x < 1$ occurring at $x_{\rm min}$ and $x_{\rm max}$, respectively. If the given $\bar b$  on the surface of the cavity falls between $b_{\rm min}$ and $b_{\rm max}$, then $\bar b = b_{q_+, \, q_-} (\bar x)$ gives three solutions $x_1 < x_2 < x_3$, which can be easily understood from, for example, Fig. 3. Only at $x_1$ or $x_3$, the corresponding slope of $b_{q_+,\, q_-} (x)$ is negative, giving the local minimal free energy. For this given pair of $(q_+, \, q_-)$, there exists a unique $b_t$ with $b_{\rm min} < b_t < b_{\rm max}$ such that the local minimal free energy at $x_1$ and that at $x_3$, with $x_1$ and $x_3$ as described above but now from $b_t = b_{q_+,\, q_-} (\bar x)$, are equal. Therefore, these two phases, one with the reduced horizon $x_1$ and the other with size $x_3$, can coexist, and the phase transition between the two is a first-order one since it involves an entropy change (note that the entropy for each phase is determined by its horizon size). One expects that like for the $D5/D1$ system, $b_t$ is a function of both $q_+$ and $q_-$ and, therefore, spans a first-order transition two-dimensional surface ending on a one-dimensional critical line,   rather than a first-order transition line ending on a critical point as for a charged black p-brane with $p < 5$. For $\bar b > b_t$, following the analysis given in \cite{Lu:2010xt}, we know that the phase with the smaller horizon size $x_1$ has the lowest free energy, therefore, the stable phase, while for $\bar b < b_t$, now the phase with the larger horizon size $x_3$ is the stable one. In other words, the smaller stable black $D6/D0$ is like the liquid phase, while the large one is like the gas phase. For a given pair of ($q_+, q_-$) with $q_- < q_+$ in the right part, for each given $\bar b$ we have a unique solution $\bar x$ from $\bar b = b_{q_+, \, q_-} (\bar x)$, and the slope of $b_{q_+, \, q_-} (x)$ at this $\bar x$ is always negative, as can be seen, for example, from Fig. 5, the corresponding free energy is lowest, and, therefore, the phase is stable.

Now the reader might wonder why adding charge to the uncharged black configuration (Schwarzschild black hole or black $p$ branes with
$p < 5$) or adding particular delocalized charged lower-dimensional branes to the original branes (for $D5$ or $D6$ branes) can modify the usual
Hawking-Page$-$type phase structure to the van der Waals$-$Maxwell liquid-gas$-$type? This is what we try to address in the next section.

\section{Origin of the phase structure modification}

One key observation for the qualitative change of the phase structure from the uncharged black configuration to the charged one
is the appearance of the divergent behavior of the reduced inverse temperature at one end $b(x \to q) \to \infty$, while the
condition at the other end $b (x \to 1) \to 0$ remains the same (note that 1 and $q$ are the upper and the lower end points of
the variable $x$, respectively). The limit $x \to q$
is actually the extremal limit, and so we can use the extremal black holes/branes to understand the reason behind the qualitative change of
phase structure, a great simplification.
\begin{figure}[htp]
\begin{center}
\includegraphics[scale=0.40, angle= -90]{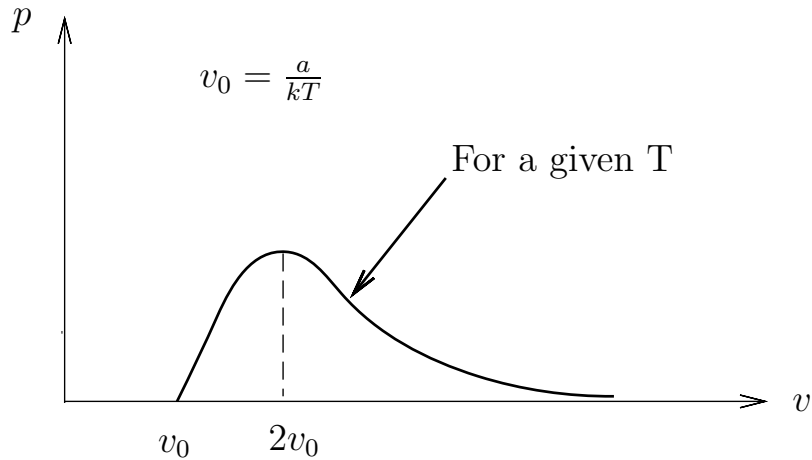}
\end{center}
\caption{The typical behavior of $p$ vs $v$ when we set $b = 0$ for the van der Waals equation of state.}
\end{figure}
Before we address the black holes/branes, let us understand the usual van der Waals liquid-gas
phase structure
described by its equation of state,\footnote{We caution the reader not to confuse the van der Waals parameters $a$ and $b$ used here with
the same parameters used for describing the $D6/D0$ system in the earlier sections.}
\be \left(p + \frac{a}{v^2}\right) (v - b) = k T,\ee
where parameter $a$ is related to the molecular attractive interaction, while $b$ is related to the repulsion. If we set $b = 0$, i.e.,
turn off the repulsive interaction, we have
\be p = \frac{k T v - a}{v^2},\ee
whose behavior is shown in Fig. 11. This is quite similar to the $b (x)$ vs $x$ diagram of uncharged black holes/branes. When we turn on the
repulsive interaction, i.e., $b \neq 0$, we have the usual van der Waals$-$Maxwell liquid-gas structure. The exact same thing happens when we add
charge to the uncharged black hole (in other words, in this case we add the repulsive interaction due to the added charge to the original
gravitational attractive interaction due to mass), giving also the van der Waals$-$Maxwell liquid-gas$-$type phase structure. This seems to suggest that
the van der Waals liquid-gas$-$type phase structure is the result of competition between the attractive and the repulsive interactions and is
independent of whether the underlying system is a liquid-gas system or a gravitational system.

This also does seem to help us understand the phase structure of the charged black $p$ brane systems. For example, adding the delocalized charged
$D(p-2)$ branes to the charged black $Dp$ branes does not change the phase structure of the original $Dp$ branes, since in the extremal limit the
interaction between the delocalized $D(p-2)$ branes and $Dp$ branes is attractive.\footnote{For interactions between branes with different dimensionalities, see, for example, \cite{polchinski-v2}.} However, adding the delocalized charged $D0$ branes to the original
$D6$ branes increases the repulsive interaction and, therefore, changes the phase structure from something similar to the chargeless case to the
van der Waals$-$Maxwell liquid-gas type. For $D5$ branes, this picture does not resolve the puzzle; namely, we know that in the extremal limit, there is
no interaction between the delocalized $D1$ branes and $D5$ branes, but the phase structure still qualitatively changes to have the van der Waals$-$Maxwell
liquid-gas type when we add delocalized $D1$ branes to $D5$ branes.

This hints at the fact that having the additional repulsive interaction is not the complete story.
In addition to providing repulsive interaction, adding charge or additional delocalized charged lower-dimensional branes can also increase the
degeneracy or the entropy of the underlying system. Note that in the canonical ensemble, the underlying phase structure is determined
by the Helmholtz free energy, which consists of two parts, the internal energy and the entropy.  Therefore, it is natural to expect that entropy also has
a role to play in addition to what has been mentioned about the nature of interactions. Let us examine in detial the origin of the divergent behavior mentioned
earlier, which is the key to the underlying phase structure.

First, let us focus on the van der Waals isotherm. We have
\bea E &=& \frac{3}{2} N k T - \frac{a N}{v}, \quad S = N k \left[\ln \frac{(v - b) T^{3/2}}{\Phi} + \frac{5}{2}\right],\nn
     p &=& -\frac{1}{N} \left(\frac{\partial F}{\partial v}\right)_{T, N} = \frac{T}{N} \left(\frac{\partial S}{\partial v}\right)_{T, N}
- \frac{1}{N}\left(\frac{\partial E}{\partial v}\right)_{T, N}\nn
     &=& \frac{ k T}{v - b} - \frac{a}{v^2},
\eea
where $F, E$, and $S$ are the free energy, internal energy, and entropy, respectively, with $F = E - TS$.
From the above, it is clear that the divergence $p \rightarrow \infty$ as $v \rightarrow b$ actually originates from the entropy. When $b = 0$,
given that $v \geq v_0 = a / k T$ (see Fig. 11), both the internal energy and the entropy of the system are finite when $v \rightarrow v_0$.
However, when we turn on $b$, i.e., $b \neq 0$, the entropy blows up when $ v \rightarrow b$, while the internal energy essentially remains
unchanged (except that we need to replace the lower end limit $v \rightarrow v_0$  by $v \rightarrow b$). In other words, the appearance of the
phase structure of van der Waals$-$Maxwell liquid gas is due to the dramatic change of entropy when $v \rightarrow b$ (with nonzero repulsive
interaction $b$). So, the repulsive core of molecules or atoms has more dramatic influence on the entropy than on the internal energy.

 Let us see what happens for the black holes. Here we have the following expressions for the so-called reduced internal energy, reduced entropy,
and reduced inverse temperature, for example, from\footnote{Our definition differs from \cite{Lundgren:2006kt} by a factor of 4.} \cite{Lundgren:2006kt}
\bea\label{rnb}
&&\tilde E = 4\left[1 - \sqrt{(1 - x) \left(1 - \frac{q^2}{x}\right)}\right],\qquad \tilde S =  x^2,\nn
&& b_q (x) = \frac{(\partial \tilde S/\partial x)_q}{(\partial \tilde E/\partial x)_q} = \frac{x (1 - x)^{1/2} \left(1 - \frac{q^2}{x}\right)^{1/2}}{1 - \frac{q^2}{x^2}}.\eea
Here we have denoted the reduced inverse temperature with a subscript $q$ to indicate that this is a charged case, and for chargeless case,
$q$ should be put to zero. It is clear from \eqn{rnb} that the divergence of $b_q (x)$ as $x \rightarrow q$ is due to the fact that $(\partial\tilde E/ \partial x)_q$
vanishes and $(\partial\tilde S/ \partial x)_q$ remains finite in this limit. This is quite different from the previous case where the divergence of $p$ was due to
the blowing up of $dS/dv$ as $v \to b$. Note that here both the entropy and the internal energy change in the same way, i.e.,
from zero in the chargeless case to a finite value in the charged case, in the respective lower end limit, i.e., $x \rightarrow 0$ (for the
chargeless case) or $x \rightarrow q$ (for the charged case).  However, their rate with respect to $x$ changes in the opposite way. For the
entropy, the rate changes from zero to a positive
finite value in the above respective lower end limit, while for the internal energy, the corresponding rate changes from a positive finite value to
zero in the same respective lower end limit.  Such a change of rate for either entropy or internal energy is due to the addition of charge
since adding charge not only gives rise to the repulsive interaction but also to the increase of degrees of freedom of the system, therefore,
the entropy. So, the vanishing of $(\partial\tilde E/ \partial x)_q$ in the limit $x \rightarrow q$ is due to the addition of charge and is mostly responsible for the blowing up of $b_q (x)$ in this same limit, therefore, for the underlying phase structure [given that the nonvanishing of $(\partial \tilde S/ \partial x)_q$ in this same limit is also important]. So, the reason for the underlying phase structure in the present case (where the rate of entropy is finite) is quite opposite
to the van der Waals isotherm (where the rate of entropy blows up) we discussed earlier.

Now, let us move on to the black $p$ brane case and see what happens there. For simple charged black $p$ branes, we have \cite{Lu:2010xt}
\bea \tilde E (x) &=& 2\left[(8 - p) - \frac{7 - p}{2}\sqrt{\triangle_+ \triangle_-} - \frac{9 - p}{2}\sqrt{\frac{\triangle_+}{\triangle_-}}\right],\nn
\tilde S (x) &=& x^{1/2} \left(1 - \frac{\triangle_+}{\triangle_-}\right)^{\frac{9 - p}{2(7 - p)}},\nn
b_q (x) &=& \frac{\left(\partial \tilde S/\partial x\right)_q}{\left(\partial \tilde E/\partial x\right)_q}
= \frac{x^{1/2}}{7 - p}\sqrt{\frac{\triangle_+}{\triangle_-}} \left(1 - \frac{\triangle_+}{\triangle_-}\right)^{\frac{p - 5}{2(7 - p)}},\eea
where $q < x <  1$ and
\be \triangle_+ = 1 - x, \qquad \triangle_- = 1 - \frac{q^2}{x}.\ee
Notice that the reduced entropy vanishes in the lower end limit either in the chargeless case ($x \rightarrow 0$) or the charged case
($x \rightarrow q$) for $p \leq 7$. Further, $(\partial\tilde S(x)/\partial x)_q$  vanishes in the chargeless case for all $p \leq 7$ in the $x \rightarrow 0$ limit, but
it blows up in the charged case only for $p < 5$, becomes a finite value for $p = 5$, and vanishes again for $p = 6$ in the extremal
limit $x \rightarrow q$. The internal energy itself changes from zero in the chargeless case to a positive finite value in the charged case in
its respective lower end limit, and $(\partial \tilde E/\partial x)_q$ is always positively finite in either case in the corresponding extremal limit for $p \leq 7$.
So, the divergent behavior of $b_q (x)$ as $x \rightarrow q$ is once again due to the blowing up of $(\partial\tilde S/\partial x)_q$  for $p < 5$, and this divergent
rate of entropy is responsible for the underlying phase structure. In other words, $p < 5$ systems behave much like the van der Waals isotherm
in the phase structure, as we discussed.

Let us consider the special case of $p = 5$. A previous study \cite{Lu:2010xt} showed that when a 5-brane is charged, the phase structure
is essentially of the same type as the chargeless case without a van der Waals$-$Maxwell liquid-gas structure, even though there are three different substructures,
analogous to the $p < 5$ cases. Further study \cite{Lu:2012rm} demonstrated that this phase structure can be qualitatively modified to a van der
Waals$-$Maxwell liquid-gas type by adding delocalized charged $D1$ branes to the black charged $D5$ branes. As discussed previously, since in the extremal
limit $x \rightarrow q_5$, there is no interaction between $D1$ branes and $D5$ branes, the divergent behavior of $b_{q_1, q_5} (x)$ must come from the blowing
up of $(\partial \bar S/ \partial  x)_{q_1, q_5} $ in this limit.  This can be understood as the addition of delocalized charged $D1$ branes increases the degeneracy of the underlying system,
therefore, the entropy. Let us examine in detial to see if this is, indeed, the case. For the $D1/D5$ system, we have\cite{Lu:2012rm}
\bea \tilde E (x) &=& 2\left[3 + \sqrt{\frac{\triangle_+}{\triangle_-}}\left(1 - G_1^{-1} \right) - 2\sqrt{ \frac{\triangle_+}{\triangle_-}}
- \sqrt{\triangle_+ \triangle_-}\right],\nn
\tilde S (x) &=& x^{1/2} \left(1 - \frac{\triangle_+}{\triangle_-}\right)\left[1 + \frac{1 - G_1^{-1}}{\frac{\triangle_-}{\triangle_+} - 1}\right]^{1/2},  \nn
b_{q_1,q_5} (x) &=&\frac{\left(\partial\bar S/\partial x\right)_{q_1, q_5}}{\left(\partial\bar E/\partial x\right)_{q_1, q_5}}
= \frac{x^{1/2}}{2}\left(\frac{\triangle_+}{\triangle_-}\right)^{1/2} \left[1 + \frac{1 - G_1^{-1}}{\frac{\triangle_-}{\triangle_+} - 1}\right]^{1/2}, \eea
where now $q_5 < x < 1$,  and
\bea \frac{\triangle_+}{\triangle_-} &=& \frac{1 - x}{1 - q_5^2/x}, \nn
 1 - G_1^{-1} &=& \frac{1}{2}\left[\sqrt{\left(\frac{\triangle_-}{\triangle_+} - 1\right)^2 + 4 q_1^2\frac{\triangle_-}{\triangle_+}} -
\left(\frac{\triangle_-}{\triangle_+} - 1\right)\right].\eea
From the above, we have $ b_{q_1,q_5} (x) \rightarrow \infty$ as $x \rightarrow q_5$. Note that $\tilde S$ continues to vanish in the extremal limit $x
\rightarrow q_5$ but $(\partial \tilde S/\partial  x)_{q_1, q_5} $ blows up in the same limit. Both the reduced internal energy $\tilde E$ and its rate $(\partial \tilde E/\partial x)_{q_1, q_5}$ are nonzero finite
in the same limit. So, the divergent behavior of $b_{q_1, q_5} (x)$ in the limit $x \rightarrow q_5$ is, indeed, due to the blowing up of $(\partial \tilde S/\partial  x)_{q_1, q_5}$ in the same limit, as anticipated.

Finally, let us consider the special case of $p = 6$. As shown in \cite{Lu:2010xt}, when charge is added to black $D6$ branes, the resulting phase
structure of charged  black $D6$ branes remains the same as its chargeless counterpart (except that we need to replace the zero of the lower end of $x$
by finite $q$). It was also shown in \cite{Lu:2012rm} and discussed in \cite{Lu:2013nt}
as well as in the previous sections in this paper that this phase structure cannot be modified to the van der Waals$-$Maxwell liquid-gas type by adding
either delocalized charged $D4$ or $D2$ branes except by adding the delocalized charged $D0$ branes. We demonstrated in the previous sections that the
phase structure for a
general $D6/D0$ system is essentially the same as that of the special case when $D0$ brane charge $Q$ is set equal to $D6$ brane
charge $P$ \cite{Lu:2013nt}.
For this reason, for simplicity, we, in what follows, just use this special case to uncover the reason behind such a change of phase structure. For the $D6/D0$
system with $q_0 = q_6 = q$ (here we are using the reduced charges of $D0$ and $D6$ branes), we have
\bea \tilde E (x) &=& 4 \left[1 - \sqrt{(1 - x)\left(1 - \frac{q^2}{x}\right)}\right],\nn
      \tilde S (x)&=& x^2, \quad (q < x < 1)\nn
      b_q (x) &=& \frac{\left(\partial\bar S/\partial x\right)_{ q}}{\left(\partial\bar E/\partial x\right)_{q}}
= \frac{x (1 - x)^{1/2}\left(1 - \frac{q^2}{x}\right)^{1/2}}{1 - \frac{q^2}{x^2}},\eea
      where now $b_q (x) \rightarrow \infty$ as $x \rightarrow q$. If we compare this case with the charged black hole discussed earlier in \eqn{rnb},
we find that we have exactly the same $\tilde E, \tilde S, b_q (x)$ in both cases. This is not surprising since it is well known that when we dimensionally reduce
this system to $D = 4$, we end up precisely with the $D = 4$ charged black hole. So, we expect that the discussion given there applies here, too. In other
words, the qualitative change of phase structure is due to the added ``repulsive interaction''. The deep reason behind this can also be understood from string/M theory
since we know that the interaction between $D0$ and $D6$ branes is repulsive, and adding delocalized $D0$ branes to the charged black $D6$ branes precisely adds this
repulsive interaction to the system making the qualitative change of phase structure possible.

With the above analysis, we understand the underlying reason for the appearance of van der Waals$-$Maxwell liquid-gas$-$type phase structure in various cases.
The key to this is to scrutinize what causes the divergent behavior of the local function, the inverse temperature $b_q(x)$, for the various black
systems in the extremal limit $x \to q$. Since we consider a canonical ensemble, the thermodynamical function of interest is the Helmholtz free energy $F =  E(x)
-  T S(x)$, where $E(x)$ and $S(x)$ are the internal energy, and the entropy, and $T$ is the preset temperature of the cavity. So, it is the rate of change of entropy
and the internal energy with respect to $x$ which are responsible for the divergent behavior of $b_q(x)$ in the extremal limit $x \to q$ and not the entropy
and internal energy themselves. When $q=0$, $E(x)$, $S(x)$, and $b_q(x)$ all vanish in the limit $x \to 0$. This has to be true given the physical context
of chargeless black system. For $q=0$, the black system has just mass, and, therefore, the interaction is only attractive. In the string/M theory context, we know
that the system has an equal number of branes and antibranes, and the net interaction has to be attractive. However, when a nonzero charge $q$ is added, actually
two ingredients are added to the system: one is the repulsive interaction (in addition to the already existing attractive one due to mass), and the other is
the increase in the degeneracy, therefore, the entropy (since adding charge is to add additional degrees of freedom). This is particularly obvious in the
context of string/M theory. These two new ingredients brought to the system when charge is added are needed for modifying the phase structure, since
the phase structure is determined by the free energy or, in turn, by the internal energy and the entropy. In string/M theory, since there exist various kinds
of branes, there are various ways to add these two ingredients to the already existing system. So, for example, we can add charges to the chargeless branes
to provide both the repulsive interaction and the additional entropy or add different kind of branes to provide more repulsive interaction (as in the case of
adding $D0$ branes to $D6$ branes) or add different kind of branes to increase the entropy (as in the case of adding $D1$ branes to $D5$ branes) of the system. This
is precisely what we have tried and succeeded for $D5$ and $D6$ branes. The addition of particular delocalized branes makes $b_q(x)$ divergent
in the $x \to q$ limit by either the blowing up of $dS(x)/dx$, or making $dE(x)/dx$ vanish, or both.

In the above, we have provided reasons for the appearance of the universal phase structure of the van der Waals$-$Maxwell liquid-gas type for various systems. This
universal phase structure is also shared by the charged AdS black hole, and in the corresponding field theory, it has similarities with the so-called catastrophic
holography \cite{Chamblin:1999tk}.  This universal phase structure is clearly the result
of the boundary condition rather than the precise details of
asymptotic metrics which can be either flat, AdS, or dS
\cite{Carlip:2003ne, Lundgren:2006kt}. The boundary condition
realized in each case by the reflecting wall actually provides a
confinement to the underlying system. This may suggest that the AdS
holography is a result of such confinement rather than the detail
properties of the AdS space. Then the natural speculation is that a similar holography should hold even in asymptotically flat space. If, indeed, such a holography holds, the
natural and interesting questions are how do we define the
corresponding field theory on the underlying holographic
screen (which is supposed to be the spherical cavity in the present case), and what do the various thermodynamical phase transitions correspond to
in the field theory so defined?

For an asymptotically flat black hole without an
origin from branes in string/M theory, establishing such a field
theory description will be extremely difficult, not to mention the
issue associated with the cavity. However, for asymptotically flat black
branes,  it is very natural to suppose that
there exists an associated dual field theory arising on the
world volume of the corresponding branes. Here one of the issues is how to
properly consider the cavity effect, which may be viewed as imposing
certain boundary conditions on the fields. Note that such a field
theory, if it exists at all, is neither supersymmetric nor conformal in
general, due to the presence of a cavity.  If we presume such a holography for the $D6/D0$ system considered,\footnote{We thank the anonymous referee for encouraging us to give
such a discussion.} the phase structure and the related properties of the charged black $D6/D0$ system placed in a cavity in a canonical ensemble are related,
by the holographic map, to the physics of the (6 + 1)-dimensional dual field theory with its fields satisfying the proper boundary conditions (which are not
clear to us at present).  For example, for each given $q_-$ with $q_- < q_+$ on the left side of the critical line given in Fig. 9, the phases
are also controlled, just like the AdS cases\cite{Chamblin:1999tk}, by the universal ``swallowtail" shapes familiar from the catastrophe theory. However,
unlike the nondilatonic (or conformal) cases and certain dilatonic (or nonconformal) cases (i.e., the $Dp$ brane cases with $p \le 4$) \cite{Klebanov:1996un,
Susskind:1998dq, Klebanov:1997kv, Peet:1998wn}, we do not have dual field theory interpretations for the entropy and free energy at a temperature
for the $D6/D0$ system.  Since the world-volume theory of the present system is related to the (6 + 1)-dimensional gauge theory, one thing is clear to us
that the $D0$ brane charge for the system is related to the condensate $< \int Tr F\wedge F \wedge F > \ne 0$ on the field theory side while keeping
$<\int Tr F> = 0, < \int Tr F \wedge F> = 0$ \cite{Taylor:1997ay}. We could say more on field theory for the $D6/D0$ system, for example, along a similar line as
for the AdS cases following \cite{Chamblin:1999tk, Chamblin:1999hg}.

However, before we embark on such discussions, we must be cautious whether the corresponding dual field theory exists at all for the present case. It is well known
that there is
no decoupling limit for the $D6$ brane theory \cite{Sen:1997we, Seiberg:1997ad, Itzhaki:1998dd, Peet:1998wn, Itzhaki:1998ka}. Actually, the $D6$ brane theory itself
is as complicated as the M theory, and for any $N$ (with $N$ the number of $D6$ branes), it is described in the UV by M theory on a flat background with $A_{N - 1}$
singularity. Note that there is no (6 + 1)-dimensional field theory in the UV (in fact, such a theory does not exist without gravity
\cite{Sen:1997we, Seiberg:1997ad}) which can flow, in the IR, to super Yang-Mills [the $D6$ theory itself flows in the IR to the (6 + 1)-dimensional super Yang-Mills]. In
\cite{Peet:1998wn}, it was also argued from the $D6$ brane low energy Hilbert space based on the result from \cite{Itzhaki:1998dd} that, most likely, there
is no underlying field theory. Adding $D0$ branes is not expected to change the situation given the undecoupled interaction of massless states from both
$D0$ and $D6$ systems \cite{Sen:1997we}.

So, the important lesson we learn from this study on the universal thermodynamical phase structure for the present $D6/D0$ system is as follows: adding the delocalized
$D0$ branes to the $D6$ brane system changes its phase structure dramatically to a very rich one exhibiting the universal feature of van der Waals$-$Maxwell liquid-gas type
as all the other brane systems ($Dp$-branes with $p\leq 4$ and $D5/D1$ system) studied previously. However, this merely reflects the thermodynamical properties of the
system in its valid description region and has its own interest. This particular system, unlike the others for which the corresponding dual field theory might exist,
does indicate that
uncovering a universal thermodynamical phase structure does not necessarily imply the existence of a holography, since for the present case the underlying
field theory does not exist as indicated in the previous paragraph. In other words, a universal thermal property and a general holography may not be necessarily
related to each other. On the other hand, if there is a holography, one should expect to see the same feature on both sides. In our discussion of the origin for
the universal phase structure for different systems in this section, we do see the difference between the $D6/D0$ system and all the other brane systems (i.e., the $Dp$
systems for $p\le 4$ and $D5/D1$ system), and we do not know if such a difference plays a role for the existence of a dual field theory description. For the former, we do not
have a dual description, but for the latter, the corresponding dual description for each case might exist, since at least the near-horizon geometry of the
corresponding system in the usual case has a dual description.  Exploration of this and the related issues will be our future program and is beyond the scope
of the present paper.

\section{Conclusion}

To conclude, in this paper we have studied the charged black $D6/D0$ bound state configuration of type IIA supergravity and its thermodynamic phase
structure with all generality. The phase structure of the same system has been studied before but only in a special case when the charges associated
with $D6$ branes and $D0$ branes are equal and that associated with the dilaton is zero. But here we have considered all the parameters of the solution to
take generic values. In general, the solution is characterized by three independent parameters. We have argued that the solution is not well defined in
the entire parameter space. There are naked singularities in a certain region of the parameter space. We have given general arguments to show that
when we restrict ourselves to a certain other region of the parameter space, then only the $D6/D0$ solution has a well defined horizon and is suitable for
studying thermodynamics. We have studied the equilibrium thermodynamics and the phase structure of the general black $D6/D0$ solution in the canonical
ensemble. For this purpose, we have computed the Euclidean action, the form of the so-called reduced inverse temperature in a suitable coordinate and expressed this
inverse temperature
in terms of a single parameter $x$ (the reduced horizon radius of the black $D6/D0$ solution). We argued that the phase structure, which is governed
by the singularity structure of the reduced inverse temperature as $x \to q$, is similar to the special case studied before. But here the analysis is much
more involved, and the phase structure is richer than that of the special case. This shows that it is a general feature (not a consequence of the
special case) that when charged delocalized $D0$ branes are added to charged $D6$ branes, the phase structure of $D6$ branes gets qualitatively changed and
takes the universal form (as for other $Dp$ branes with $p<5$) which has van der Waals$-$Maxwell liquid-gas$-$type structure. We have tried to unravel the
reasons why such a drastic change in phase structure occurs when charges and/or other branes are added to the existing system. We have shown in
a case-by-case basis that adding charge and/or other branes actually adds either the repulsive interaction or the additional degrees of freedom, i.e.,
entropy to the system. These two ingredients are actually causing the qualitative change of phase structure to the universal form in various cases.

\section*{Acknowledgements}

J.X.L and J.O. acknowledge support from the NSF of China with Grant No. 11235010.


\begin{thebibliography}{99}
\bibitem{Chamblin:1999tk}
  A.~Chamblin, R.~Emparan, C.~V.~Johnson, and R.~C.~Myers,
  Phys.\ Rev.\  D {\bf 60}, 064018 (1999).

\bibitem{Chamblin:1999hg}
A.~Chamblin, R.~Emparan, C.~V.~Johnson, and R.~C.~Myers,
  Phys.\ Rev.\  D {\bf 60}, 104026 (1999).

\bibitem{Kubiznak:2012wp}
   D.~Kubiznak and R.~B.~Mann,
   J. High Energy Phys. {\bf 07} (2012) 033.


\bibitem{Gunasekaran:2012dq}
   S.~Gunasekaran, R.~B.~Mann, and D.~Kubiznak,
   J. High Energy Phys. {\bf 11} (2012) 110.


\bibitem{Altamirano:2013uqa}
   N.~Altamirano, D.~Kubiz¨¾¨¢k, R.~B.~Mann, and Z.~Sherkatghanad,
   Classical Quantum Gravity  {\bf 31}, 042001 (2014).


\bibitem{Smailagic:2012cu}
    A.~Smailagic and E.~Spallucci,
   Int.\ J.\ Mod.\ Phys.\ D {\bf 22}, 1350010 (2013).

 \bibitem{Spallucci:2013osa}
    E.~Spallucci and A.~Smailagic,
   Phys.\ Lett.\ B {\bf 723}, 436 (2013).

\bibitem{Spallucci:2013jja}
    E.~Spallucci and A.~Smailagic,
   J.\ Gravit.\  {\bf 2013}, 525696 (2013).

\bibitem{Nicolini:2011dp}
   P.~Nicolini and G.~Torrieri,
   J. High Energy Phys. {\bf 08} (2011) 097.


\bibitem{Hawking:1982dh}
  S.~W.~Hawking and D.~N.~Page,
  Commun.\ Math.\ Phys.\  {\bf 87}, 577 (1983).

\bibitem{Witten:1998zw}
  E.~Witten,
  Adv.\ Theor.\ Math.\ Phys.\  {\bf 2}, 505 (1998).

\bibitem{Carlip:2003ne}
  S.~Carlip and S.~Vaidya,
  Classical Quantum Gravity  {\bf 20}, 3827 (2003).

\bibitem{Lundgren:2006kt}
  A.~P.~Lundgren,
  Phys.\ Rev.\  D {\bf 77}, 044014 (2008).


\bibitem{Lu:2010xt}
  J.~X.~Lu, S.~Roy, and Z.~Xiao,
  J. High Energy Phys. {\bf 01} (2011) 133.


\bibitem{Lu:2012rm}
  J.~X.~Lu, R.~Wei, and J.~Xu,
 J. High Energy Phys. {\bf 12} (2012) 012.

\bibitem{Lu:2013nt}
  J.~X.~Lu and R.~Wei,
  J. High Energy Phys. {\bf 04} (2013) 100.


\bibitem{Brandhuber:1997tt}
  A.~Brandhuber, N.~Itzhaki, J.~Sonnenschein, and S.~Yankielowicz,
  Phys.\ Lett.\ B {\bf 423}, 238 (1998).

\bibitem{Dhar:1998ip}
  A.~Dhar and G.~Mandal,
   Nucl.\ Phys.\ {\bf B531}, 256 (1998).


\bibitem{York:1986it}
  J.~W.~York,
  Phys.\ Rev.\  D {\bf 33}, 2092 (1986).

\bibitem{Gibbons:1976ue}
  G.~W.~Gibbons and S.~W.~Hawking,
  Phys.\ Rev.\  D {\bf 15}, 2752 (1977).

\bibitem{polchinski-v2}
J.~ Polchinski, Superstring Theory (Cambridge University Press, Cambridge, Enland, 1998), Vol. 2.

\bibitem{Taylor:1997ay}
  W.~Taylor,
   Nucl.\ Phys.\ {\bf B508}, 122 (1997).

\bibitem{Klebanov:1996un}
  I.~R.~Klebanov and A.~A.~Tseytlin,
  Nucl.\ Phys.\ {\bf B475}, 164 (1996).

\bibitem{Susskind:1998dq}
  L.~Susskind and E.~Witten,
  arXiv:hep-th/9805114.

\bibitem{Klebanov:1997kv}
  I.~R.~Klebanov and L.~Susskind,
  Phys.\ Lett.\ B {\bf 416}, 62 (1998).

\bibitem{Peet:1998wn}
  A.~W.~Peet and J.~Polchinski,
  Phys.\ Rev.\ D {\bf 59}, 065011 (1999).

\bibitem{Sen:1997we}
  A.~Sen,
  Adv.\ Theor.\ Math.\ Phys.\  {\bf 2}, 51 (1998).

\bibitem{Seiberg:1997ad}
  N.~Seiberg,
  Phys.\ Rev.\ Lett.\  {\bf 79}, 3577 (1997).

\bibitem{Itzhaki:1998dd}
  N.~Itzhaki, J.~M.~Maldacena, J.~Sonnenschein, and S.~Yankielowicz,
  Phys.\ Rev.\ D {\bf 58}, 046004 (1998).



\bibitem{Itzhaki:1998ka}
  N.~Itzhaki,
  J. High Energy Phys. {\bf 09} (1998) 018.






\end{thebibliography}
\end{document}